\documentclass[10pt]{article}

\input epsf

\usepackage{amsmath}
\usepackage{amscd}
\usepackage{amsfonts,amssymb}
\usepackage{epsfig}

\newcommand{\news}{\setcounter{equation}{0}}
\newcommand{\be}{\begin{equation}}
\newcommand{\ee}{\end{equation}}
\newcommand{\bea}{\begin{eqnarray}}
\newcommand{\eea}{\end{eqnarray}}

\newcommand{\bal}{\begin{align}}
\newcommand{\eal}{\end{align}}

\newcommand{\bean}{\begin{eqnarray*}}
\newcommand{\eean}{\end{eqnarray*}}
\font\upright=cmu10 scaled\magstep1

\newcommand{\Z}{\mathbb{Z}}

\newcommand{\C}{\mathbb{C}}
\newcommand{\R}{\mathbb{R}}

\newcommand{\identity}{{\upright\rlap{1}\kern 2.0pt 1}}
\newcommand{\half}{\frac{1}{2}}
\newcommand{\pr}{\partial}

\newcommand{\M}{{\cal M}}

\newcommand{\K}{{\cal K}}

\newcommand{\bz}{\bar z}
\newcommand{\bZ}{\bar Z}

\begin{document}
\pagestyle{plain}
\title{\vskip -70pt
\begin{flushright}
{\normalsize DAMTP-2010-63} \\
\end{flushright}
\vskip 30pt
{\bf \LARGE Vortices and Jacobian Varieties}
 \vskip 30pt}
\author{Nicholas S.\ Manton\thanks{\tt N.S.Manton@damtp.cam.ac.uk} 
\,\,\,and\,\,\,Nuno M.\ Rom\~ao\thanks{\tt nromao@imf.au.dk} \\ \\
{\sl ${}^*$ Department of Applied Mathematics and Theoretical Physics,}\\
{\sl University of Cambridge,}\\
{\sl Wilberforce Road, Cambridge CB3 0WA, England}\\[12pt]
{\sl ${}^\dagger$ Centre for Quantum Geometry of Moduli Spaces,}\\
{\sl Institute of Mathematical Sciences, University of Aarhus,}\\
{\sl Ny Munkegade bygn.\ 1530, 8000 \AA rhus C, Denmark} \\[12 pt]
{\sl ${}^\dagger$ Institute of Mathematics}\\
{\sl Jagiellonian University, Cracow}\\
{\sl ul.\ \L ojasiewicza 6, 30-348 Krak\'ow, Poland}\\[12pt]
}

\date{October 2010}
\maketitle

\begin{abstract}
We investigate the geometry of the moduli space of  $N$-vortices on
line bundles over a closed Riemann surface $\Sigma$ of genus $g> 1$, in 
the little explored situation where $1\le N<g$. In the regime where 
the area of the surface is just large enough to accommodate $N$
vortices (which we call the dissolving limit), we describe the
relation between the geometry of the moduli space and the complex 
geometry of the Jacobian variety of $\Sigma$. For $N=1$, we show that the 
metric on the moduli space converges to a natural Bergman metric
on $\Sigma$. When $N>1$, the vortex metric typically degenerates as the dissolving 
limit is approached, the degeneration occurring precisely on the
critical locus of the Abel--Jacobi map of $\Sigma$ at degree $N$. We 
describe consequences of this phenomenon from the point of view of 
multivortex dynamics. 
\vspace{50pt}

\end{abstract}

\vskip 80pt

\newpage
\section{Introduction}
\news

Inspired at first by the Ginzburg--Landau theory of superconductivity, 
several models for the dynamics of magnetic flux vortices 
\cite{JafTau, Dun, Yang,ManSut} have been studied as gauge theories in $2+1$ 
dimensions. The surface supporting a vortex is usually considered as a 
cross section of a ``true'' vortex tube in three spatial dimensions,
but the models often make sense purely 
in two spatial dimensions, on surfaces like the euclidean and the 
hyperbolic planes, with boundaries at infinity, and also on closed
surfaces such as the 2-sphere. In the present paper, we will be focusing on 
what is known as the abelian Higgs model for vortex dynamics, on a closed
surface $\Sigma$ of genus $g>1$. In this particular setting, several 
interesting results have already been gathered 
in the literature, some of them in analogy with the most extensively 
studied case where $\Sigma$ is taken to be the euclidean plane.

The time-independent model consists of a system of coupled partial
differential equations 
for a connection ${\rm d}_a$ and a section $\phi$ of a 
complex line bundle  ${\mathcal L} \rightarrow \Sigma$. On the
surface $\Sigma$, a K\"ahler structure $(\Sigma, j_\Sigma, \omega_\Sigma)$, where $j_\Sigma$ is
the complex structure and $\omega_\Sigma$ the K\"ahler 2-form (here 
just the area form) needs to be 
specified, and the bundle $\mathcal L$, supposed to be nontrivial, is 
equipped with a hermitian metric. A real parameter $\tau$ appears 
in the energy and in the field equations. This determines the size of a
vortex, which is significant when the geometry of $\Sigma$ is
fixed. Another parameter, the coupling constant $\lambda$, will be
fixed at $\lambda=1$. This models the boundary between type I and 
type II superconductors, and turns out to be most interesting 
mathematically, since the field configurations of minimal energy
then satisfy the first-order {vortex equations} \cite{Bog}. Here there 
are no static forces between vortices; for generic allowed values of 
the parameter $\tau$,  a unique static multivortex solution, up to 
gauge transformations, has been proved to exist for any distribution 
of $N$ points on $\Sigma$~\cite{Brad,Gar}. These points specify isolated zeroes of the 
section $\phi$. The positive integer $N$, the vortex number, 
is also the total magnetic flux through the
surface in units of the ``classically quantised'' flux of a single 
vortex. Although the vortices should be regarded as extended 
objects over the surface, these points on $\Sigma$ (which can be
superposed) can be seen as precise locations of $N$ vortex centres,
around which the energy density of the fields typically concentrates. 
In this regime, the moduli space of $N$-vortex solutions is thus the 
$N$th symmetric power of the surface $\Sigma$,
\be
{\cal M}_N \cong {\rm Sym}^N \Sigma := \Sigma^N /\mathfrak{S}_N
\ee
where $\mathfrak{S}_N$ is the symmetric group, and is a 
smooth complex manifold. In particular $\M_1$, as a complex manifold, 
is just a copy of $\Sigma$. 

The dynamical model extends the time-independent model by including a
kinetic energy term in the lagrangian. This is quadratic in the time
derivatives of the fields. Also the connection needs to be extended to
include a time-component. Since there is no static potential 
interaction between vortices when $\lambda =1$, the non-relativistic motion 
of $N$ vortices on $\Sigma$ reduces to a geodesic motion on the moduli 
space $\M_N$, with respect to a Riemannian metric
(for each value of $\tau$) that arises from the kinetic part of the 
lagrangian \cite{Man1,Stu}. This metric can alternatively be thought 
of as being induced from the  natural $L^2$-metric on the space of 
solutions of the vortex equations. 

Quite a lot is known about this family of $L^2$-metrics on the 
moduli space $\M_N$ parametrised by $\tau$. A very interesting
result is a formula due to Samols expressing the 
metric in terms of local data 
around the vortex centres ~\cite{Sam}. Each metric in the family is K\"ahler
with respect to the natural complex structure on ${\rm Sym}^N \Sigma$
induced from the original one on $\Sigma$, which is independent of $\tau$. From 
Samols' formula, one can derive an explicit expression for 
the total volume of $\M_N$ \cite{ManNas}. This
volume depends only on the topological data (the genus $g$ of $\Sigma$ 
and the vortex number $N$), on the total area $A=\int_\Sigma \omega_\Sigma$ 
of $\Sigma$, and on $\tau$. It does 
not depend on more detailed metric information. 

The allowed range of the parameter
$\tau$ for solutions of the vortex equations to exist is the 
interval $[\frac{4 \pi N}{A},+\infty [$, for a given $N>0$.
In the present work, we will be interested in the situation 
where $\tau$ approaches the lower end of this interval. Specifically, 
at $\tau=\frac{4 \pi N}{A}$ the vortex equations 
have solutions, but $\phi$ vanishes everywhere --- so, strictly 
speaking, there are no vortices. There remains a magnetic field,
which is a multiple of the area form on the surface. There also 
remain moduli, because the magnetic field does not completely fix 
the holonomies (the Wilson loop variables) of the connection around 
noncontractible 1-cycles on $\Sigma$. The moduli space of connections in 
this limit of ``dissolved vortices'' is in fact a 
translate of the moduli space of flat ${\rm U}(1)$-connections on $\Sigma$. 
This is the dual of the Jacobian variety ${\rm Jac} (\Sigma)$ associated 
to $\Sigma$, a torus of complex dimension $g$, for any value of $N$. 
Importantly, if the holonomies change 
with time, there will be electric fields on $\Sigma$ and hence a positive 
field kinetic energy. The moduli space in this regime of dissolved
vortices, which we can 
identify with the Jacobian itself by duality, therefore
acquires a metric. This metric has been shown by Nasir to be 
flat~\cite{Nas}, and it only depends on the complex structure of $\Sigma$,
not on the detailed form of the conformal factor $\Omega$. We 
rederive these results below.

Close to this critical situation, in the case where $\tau$ is
arbitrarily close to but greater than 
$\frac{4\pi N}{A}$, the moduli space $\M_N$ has complex dimension 
$N$. The picture is that the section $\phi$ is close to zero 
everywhere (as its $L^2$-norm equals $\tau A - 4 \pi N$ for static fields),
but its zero locus consists of isolated points.  In this paper, we 
shall describe an approximation to the vortex equations modelling 
this situation, and will call it the regime of ``dissolv{\em ing} vortices''
(as opposed to dissolv{\em ed}, which we defined in the preceding paragraph).
Specifically, we will be mainly dealing with the  case where $N <
g$. (Dissolving vortices in the cases where $g=1$ and $N=1$, and where $g=0$ 
with $N$ arbitrary, were investigated previously in 
\cite{Man2} and \cite{BapMan}, respectively.) 
The moduli space $\M_N$ can be identified with the space of effective
divisors of order $N$ on the algebraic curve $\Sigma$, i.e.\ sets of $N$ (not 
necessarily distinct) points on $\Sigma$ --- the zeros of $\phi$ with 
multiplicities counted algebraically. 
In algebraic geometry, any divisor on $\Sigma$ gives rise to a holomorphic line
bundle, but as $\phi$ has no poles, we need to restrict to the
effective divisors, which give rise to line bundles with nontrivial
holomorphic sections. An effective divisor is converted to a
point in the Jacobian by the Abel--Jacobi map \cite{GriHar,Jost}. 
At generic points of the image, the Abel--Jacobi map embeds $\M_N$ 
smoothly (and holomorphically) in
${\rm Jac}(\Sigma)$, but there can be singular points. However, the manifold $\M_1$ 
of effective divisors of order 1 is simply $\Sigma$, and the Abel--Jacobi 
map embeds $\M_1$ smoothly in ${\rm Jac}(\Sigma)$.

Our main result for the $N=1$ case is the following \\[5pt]
{\bf Theorem 1:} {\em The metric describing the motion of one
dissolving vortex is a natural Bergman
metric on $\Sigma \cong {\cal M}_1$. It coincides with the K\"ahler 
metric obtained by pulling back the flat metric dual to the metric 
of dissolved vortices (which is determined by the polarisation of 
the Jacobian), via the Abel--Jacobi embedding 
${\rm AJ}_1:\Sigma  \hookrightarrow {\rm Jac}(\Sigma)$.} \\[5pt]
The proof of this Theorem is contained in Section~\ref{1vortex}.
(The notion of Bergman metric we are using is explained in Appendix A, 
and most of the other ingredients in Section~\ref{geomJac}.)
As $\tau$ grows away from $\frac{4\pi}{A}$, the metric on ${\cal M}_1$
is a deformation of this Bergman metric defined from the 
complex structure alone, and
this deformation will incorporate details of the metric data on $\Sigma$ (Riemannian
metric and hermitian structure) needed to write down the lagrangian of
the model. Unfortunately, the metric on $\M_1$ is not known explicitly except 
in limiting cases. However, $\M_1$ is always in the same conformal
class as $\Sigma$, and its total area is an affine function
of both $\tau A$ and $g$. The opposite to the dissolving limit is the 
regime where $\tau \rightarrow + \infty$, which has been investigated 
previously~\cite{Man2}. Here the local geometry near the vortex is 
crucial. The metric at a point of $\M_1$ is a constant multiple of 
the metric on $\Sigma$ with a leading correction 
that depends on the curvature of $\Sigma$ at that point. The interpretation is
that the vortex behaves almost exactly like a localised
particle on $\Sigma$. So the flow of the one-vortex moduli space metric (as $\tau$
increases) turns an effective divisor, which is not a localised
object when the associated holomorphic bundle and a specific 
holomorphic section solving the vortex equations are considered, 
into a localised particle.

Broadly speaking, when $N>1$, the slow motion of a vortex is determined by two
factors: on the one hand, the nature of the metric on $\Sigma$ close to the
vortex and the hermitian structure on the bundle 
${\mathcal L}\rightarrow \Sigma$, which are essentially
introduced by hand in the model, and on the other hand, the 
effect of other nearby vortices, which in turn is controlled by the 
parameter $\tau$ scaling the area $\frac{4 \pi}{\tau}$ of an 
``effective disc'' occupied by each vortex.
The most salient feature of the latter effect is that the
head-on symmetric scattering of two vortices along some curve of $\Sigma$ occurs 
in finite time, with the vortices emerging along
a perpendicular curve on $\Sigma$ through the point of collision. 
This conclusion follows essentially from the fact that
the directions of geodesics for a K\"ahler metric are determined 
by the complex structure alone; the simplest version of this 
local argument wams first spelled out by Hitchin in~\cite{HitGTMS}. 
For dissolving vortices on a round sphere, the behaviour of vortex
scattering geodesics more general than head-on was described by 
Baptista and Manton~\cite{BapMan} --- however, on a round sphere the metric 
has positive curvature, and the general scattering behaviour of
dissolving vortices on a surface $\Sigma$ of higher genus will be quite different.

In Section~\ref{mvortices}, we discuss the geometry of the slow motion
of dissolving multivortices, in the case that $1<N<g$. It is natural 
to expect that the Abel--Jacobi map
\begin{equation} \label{AJ}
{\rm AJ}_N:{\rm Sym}^N \Sigma \longrightarrow {\rm Jac (\Sigma)}
\end{equation}
still relates the geometry of dissolving vortices with that of 
dissolved $N$-vortices, but it turns out that
this map typically has singular points (more precisely, there are 
singular points unless $N=2$, $g=3$ and
$\Sigma$ is not hyperelliptic). So the pull-back of the dual of the
metric of dissolved vortices degenerates at the critical locus of ${\rm AJ}_N$:
\\[5pt]
{\bf Theorem 2:} {\em For $1<N<g$, the slow motion of dissolving vortices is 
described generically by the pull-back of the dual of the flat metric
of dissolved vortices via the Abel--Jacobi map (\ref{AJ}). This
pull-back is a K\"ahler metric in an open subset of ${\cal M}_N$,
degenerating at the locus of special effective divisors of
degree $N$}.
\\[5pt]
Geodesics of these metrics of dissolving multivortices follow 
the expected pattern
of $90^{o}$ vortex scattering, but they become rather exotic at the
degeneration locus, which consists of the union of exceptional fibres of the
Abel--Jacobi map. The complement of the set of special divisors 
consists of the points around which the Abel--Jacobi map is a 
one-to-one cover. To understand what happens at the boundary 
of this regularity locus, we study the simplest case in
Section~\ref{singular}, namely, the behaviour of two dissolving 
vortices on a hyperelliptic Riemann surface of genus 3 near 
the singularity. Then the image of the Abel--Jacobi map is a 
surface $W_2\subset {\rm Jac} (\Sigma)$ with a double point, and the 
moduli space ${\cal M}_2 = {\rm Sym}^2 \Sigma$ can be recovered from 
it by blowing up this singularity. We will show that the geodesic 
motion will be suppressed along the directions of the exceptional fibre
introduced in the blow-up.
As we shall explain in Section~\ref{singular}, an interesting
implication of this observation is that the motion
of two dissolving vortices on the surface is able to 
detect Weierstra\ss\ points.

\section{Vortex dynamics and the vortex equations} \label{dynamics}
\news

Let $\Sigma$ be a closed Riemann surface of genus $g$ with a compatible 
metric $g_\Sigma$. We will be working on the fixed space-time $\R \times \Sigma$ 
with lorentzian metric
\be
{\rm d}s^2 = {\rm d}t^2 - g_\Sigma \,,
\ee
where $t$ is a time coordinate on $\R$. In terms of a local complex
coordinate $z$ on $\Sigma$, $g_\Sigma = \Omega \,{\rm d}z {\rm d}\bz$, 
where $\Omega$ is a positive real function called the conformal factor, 
and the associated K\"ahler 2-form is
$\omega_\Sigma=\frac{\rm i}{2}\Omega\,{\rm d}z\wedge{\rm d}\bz$.

Over $\Sigma$, we also fix a hermitian line bundle 
${\mathcal L}\rightarrow \Sigma$, i.e.\ for each $P\in \Sigma$ the
fibre ${\mathcal L}_P$ is a copy of $\mathbb{C}$ endowed with 
a hermitian inner product $\langle\cdot,\cdot\rangle_P$ that varies 
smoothly over $\Sigma$. The hermitian inner product can be used to
normalise local trivialisations,
and this amounts to reducing the structure group of ${\mathcal L} 
\rightarrow {\Sigma}$ from $\mathbb{C}^*$ to the subgroup ${\rm U}(1)$. 
We can use the projection ${\rm pr}_\Sigma$ onto $\Sigma$ to pull $\mathcal L$ 
back to our spacetime $\R \times \Sigma$, and the resulting line bundle 
${\rm pr}_\Sigma^* {\mathcal L}  \rightarrow \mathbb{R}\times \Sigma$ can be described by
time-independent transition functions.

The dynamics we want to discuss involves configurations of two 
fields on $\R \times \Sigma$: a complex field $\phi$ which is a smooth section 
of ${\rm pr}_\Sigma^*{\mathcal L}$, and a ${\rm U}(1)$-connection 
${\rm D}_{\hat a}$ on ${\rm pr}_\Sigma^*{\mathcal L}$. On a local 
trivialisation, $\phi$ is equivalent to a complex function, whereas 
the connection can be expressed as the covariant derivative 
\be
{\rm D}_{\hat a}={\rm d}-{\rm i} {\hat a} \,,
\ee
where  ${\hat a} = a_t \, {\rm d}t + a_z \, {\rm d}z +
a_{\bz} \, {\rm d}\bz$ is a real 1-form,  with $z$ a local complex 
coordinate on the trivialising open set
$U\subset \Sigma$ and $a_{\bz} = \overline{a_z}$. 
Gauge transformations act on the fields as
\bea
\phi &\mapsto& {\rm e}^{{\rm i}\chi} \phi \label{gaugetrm1}\\
{\rm D}_{\hat a} &\mapsto& 
{\rm D}_{\hat a} -{\rm i} \,{\rm d}\chi \,, \label{gaugetrm2}
\eea
with $\chi$ a real function.

The connection splits into a time and space part, ${\hat a} = a_t \, {\rm d}t +
a$, where, locally, $a_t$ is a real function and $a$ is a real
1-form on $\Sigma$, both time-dependent in general. Then the covariant 
derivative of $\phi$ splits as
\be
{\rm D}_{\hat a}\phi = {\rm D_t}\phi \, {\rm d}t + {\rm d}_a\phi
\ee
where ${\rm D_t}\phi = \partial_t\phi -{\rm i} a_t \phi$ is a section of 
${\rm pr}^*_{\Sigma}{\mathcal L}\rightarrow \mathbb{R}\times \Sigma$ 
and ${\rm d}_a\phi = {\rm d}\phi -{\rm i} a \phi$ is a path of 1-forms on
$\Sigma$ with values in ${\mathcal L}$  (locally, these two quantities are just a complex function on space-time
and a time-dependent 1-form on $\Sigma$, respectively). 
The (space-time) Maxwell 2-form is the curvature of the connection,
$F_{\hat a} = {\rm d}{\hat a}$, which can be written as
\be \label{Max}
F_{\hat a} = {\rm d}t \wedge e_a + b_a \,,
\ee
where $e_a = \partial_t a - {\rm d}a_t$ (the electric field) is a 
time-dependent 1-form and $b_a = {\rm d}a$ (the magnetic field) 
is a time-dependent 2-form, both globally defined on $\Sigma$ and both
invariant under gauge transformations.
Complex line bundles ${\mathcal L} \rightarrow \Sigma$ are topologically 
classified by their
Chern number (or degree) $N \in \mathbb{Z}$, which is given by the integral
\be
N = \frac{1}{2\pi}\int_\Sigma b_a \;\; \in \mathbb{Z}
\label{ChernNo}
\ee
and can be interpreted as a ``quantised'' total magnetic flux.
Throughout this paper, we shall assume that $N$ is positive.

The lagrangian of the abelian Higgs model at coupling
$\lambda=1$ is~\cite{ManSut}
\be
L = \frac{1}{2}\int_\Sigma \left\{ e_a\wedge *e_a  
+ {\rm D_t}\phi\wedge*\overline{{\rm D_t}\phi}- b_a\wedge *b_a 
 -{\rm d}_a\phi\wedge*\overline{{\rm d}_a \phi}
-\frac{1}{4}*\left(\langle\phi,\phi\rangle-\tau\right)^2\right\}
\label{lagran}
\ee
where $\tau$ is a fixed positive parameter.
We are using the Hodge $*$-operator associated to $g_\Sigma$, which in
terms of the coordinate $z$ acts on forms as 
$*{\rm d}z=-{\rm i}\,{\rm d}z$, $*{\rm d}\bz={\rm i}\,{\rm d}\bz$,
$*1=\omega_\Sigma$ and $*\omega_\Sigma=1$. 
The contribution of the first two terms, involving the 
time derivatives of the fields and $a_t$, defines the kinetic energy of the
theory, $T$, and the remaining terms give (minus) the potential
energy, $V$. Note that the kinetic energy is independent of 
the parameter $\tau$. Also note that the first and fourth terms, 
involving 1-forms and their Hodge duals, 
are conformally invariant --- they are unchanged if one 
deforms $g_\Sigma$ within the same conformal
class (in other words, keeps the complex structure on $\Sigma$ fixed). 

Static vortices, associated to fields with no time dependence 
and vanishing $a_t$, minimise the potential energy
\be
V = \frac{1}{2}\int_\Sigma 
\left\{b_a\wedge*b_a+{\rm d}_a\phi\wedge*\overline{{\rm d}_a\phi}
+\frac{1}{4}*\left(\langle\phi,\phi\rangle-\tau\right)^2\right\} \,.
\ee
By a standard reorganisation of the integral, it can be shown that the minimal
value of $V$ is $\pi\tau N$, and is attained precisely when the fields satisfy
the first-order vortex equations on $\Sigma$ \cite{Bog}
\bea
&\bar\partial_a \phi = 0 \,, \label{Bogo1}&\\
&*b_a+\frac{1}{2} (\langle \phi, \phi\rangle -\tau) =0\,.&\label{Bogo2}
\eea
The first equation expresses that the section $\phi$ of 
${\mathcal L}\rightarrow \Sigma$ is holomorphic, i.e.\  annihilated by the operator 
$\bar\partial_a:\Omega^{0}(\Sigma,{\mathcal L})
\rightarrow\Omega^1(\Sigma,{\mathcal L})$ 
(locally, $\bar\partial_a=\bar\partial-{\rm i}a_{\bz}{\rm d}\bz$)
defined from the unitary connection ${\rm d}_a$ and the complex 
structure on $\Sigma$~\cite{DonKro}. The second equation relates
the curvature $b_a={\rm d}a$ of the connection to the moment map 
of the holomorphic, hamiltonian action of ${\rm U}(1)$ on the 
fibres of ${\mathcal L}\rightarrow \Sigma$ (with K\"ahler structure 
induced from the hermitian metric) evaluated after $\phi$. The 
presence of the constant $\tau \in \mathbb{R}$ relates to the 
ambiguity in the choice of a moment map for this action.
 
By integrating (\ref{Bogo2}) over $\Sigma$ and using  (\ref{ChernNo}), one finds
\be
||\phi||^2_{L^2}=\tau A - 4 \pi N\
\label{Bogo2int}
\ee
where $A:=\int_\Sigma \omega_\Sigma$ is the total area of the surface 
and $||\phi||^2_{L^2}:=\int_\Sigma \langle \phi,\phi\rangle \omega_\Sigma$. 
Since this squared $L^2$-norm is non-negative, the vortex
equations can only have solutions if $\tau \ge \frac {4\pi N}{A}$. 
If we take $\tau > \frac {4\pi N}{A}$, there is a unique 
solution, up to gauge equivalence, for any choice of $N$ unordered
(not necessarily distinct) points on $\Sigma$ where $\phi$ is required 
to vanish \cite{Brad,Gar}. This is what is called an $N$-vortex 
solution. The moduli space $\M_N$ of $N$-vortex solutions is 
therefore ${\rm Sym}^N \Sigma=\Sigma^N/{\mathfrak{S}_N}$, a smooth complex 
manifold with complex dimension $N$. 

We shall refer to the critical situation where 
$\tau = \frac{4\pi N}{A}$ as the limit of {\em dissolved} vortices. 
Here, solutions of the vortex equations also exist. Since $||\phi||^2_{L^2}$
must vanish, and $\phi$ is smooth, $\phi = 0$ everywhere, so there 
are no true (localised) vortices. The first vortex equation is now
trivially satisfied and the second vortex equation reduces to 
\be \label{projflatb}
b_a = \frac{\tau}{2} \omega_\Sigma \,,
\ee
which means that the magnetic flux per unit area 
has the constant value $\frac{2 \pi N}{A}$. Notice that the value 
of this constant is determined by the topology
and the total area $A$ of $\Sigma$.
A connection ${\rm d}_a$ satisfying equation (\ref{projflatb}) 
is called a projectively flat (or constant curvature) connection, and it
always exists, but is not completely determined by the magnetic 
field $b_a$ if $g \ge 1$ (which implies that $\Sigma$ is not 
simply connected). We shall see below that the 
space of such connections, up to gauge equivalence, is a flat
torus of real dimension $2g$, irrespective of the value of $N$.

To investigate vortices and their moduli space in what we shall 
call the {\em dissolving limit},
where $\tau$ slightly exceeds $\frac{4\pi N}{A}$, it is sufficient to make 
the approximation that $\phi$ is small, and to neglect the term 
$\langle \phi,\phi\rangle$ 
in the second vortex equation. In this regime, the connection is 
therefore taken to satisfy (\ref{projflatb}), as discussed
above; there will be corrections to this, which in principle could 
be studied in perturbation theory in the parameter 
$\varepsilon=\tau-\frac{4\pi N}{A}$, but we shall not pursue this here.
In addition, the connection must 
be such that the first vortex equation has a nontrivial 
solution. We shall see that this prescription (which substitutes 
the first vortex equation {\sl per se}) picks out a subset of the 
connections that occur in the situation of dissolved vortices.

\section{Dissolved vortices} \label{dissolved}
\news

In this section, we shall discuss the critical situation of dissolved
vortices, where
\be
\tau = \frac{4\pi N}{A} \,.
\ee
Here, the field $\phi$ vanishes everywhere on $\Sigma$, so
the first vortex equation (\ref{Bogo1}) is trivially satisfied. The second 
vortex equation (\ref{Bogo2}) fixes the magnetic field on $\Sigma$ and reduces 
to the equation for a projectively flat connection
\be \label{projflatf}
b_a = {\rm d}a = \frac{\tau}{2} \, \omega_\Sigma \,.
\ee
The metric on $\Sigma$ is K\"ahler, so locally there is a real 
K\"ahler potential $\K$ such that
\be
\omega_\Sigma= {{\rm i}} \pr \bar \pr \K \, .
\ee
Therefore, a choice for the connection 1-form is, locally,
\be
a = \frac{{\rm i} \tau}{4}( \bar\pr \K-\pr \K ) \,,
\label{connK}
\ee
which is real. The ambiguity in the local
K\"ahler potential corresponds to an ambiguity in the choice of
gauge.

Globally, there is not a unique unitary connection for this magnetic 
field. The general such connection can be expressed as 
${\rm d}_a - {\rm i}\alpha$, where $a$ is fixed as above, and 
$\alpha$ is a global, real, closed 1-form, 
satisfying ${\rm d}\alpha = 0$. If $\alpha$ is globally an exact form, then 
$a \mapsto a+\alpha$ is simply a gauge transformation of $a$. To 
project out gauge transformations, it is natural to impose the further
condition ${\rm d}*\alpha = 0$.
With this prescription, $\alpha$ (satisfying ${\rm d}\alpha 
= {\rm d}*\alpha = 0$) is a real, harmonic 1-form on $\Sigma$. 

Now let us consider a time-varying connection, with the same unchanging
magnetic field. This is described by a connection 
${\rm d}_{a + \alpha}={\rm d}_a-{\rm i}\alpha$, 
where ${\rm d}_a$ is fixed in time and as above, together with 
a time-varying, harmonic 1-form $\alpha$. In addition, there is the 
real function $a_t$, also varying in time. The
1-form electric field is $e=e_{a+\alpha} 
= \pr_t \alpha - {\rm d} a_t$, and should satisfy Gau\ss's law, 
which is ${\rm d}*e= 0$ when $\phi$ vanishes. (Gau\ss's law is one 
of the Euler--Lagrange equations arising in the abelian Higgs model, and 
expresses the constraint that time variations of a connection should 
be projected orthogonally to the orbits of the group of gauge transformations 
in the space of infinitesimal connections, with respect to the $L^2$-norm.)
With $\alpha$ harmonic, 
Gau\ss's law is satisfied by setting $a_t = 0$, and the electric field 
is simply $e= \pr_t \alpha$. The kinetic energy is then
\be
T = \half \int_\Sigma e \wedge *e = \half \int_\Sigma  (\pr_t \alpha) 
\wedge *(\pr_t \alpha) \,.
\ee

The space of real, harmonic 1-forms on $\Sigma$ has real dimension $2g$,
and is isomorphic to the space 
of holomorphic 1-forms on $\Sigma$, with
complex dimension $g$.
This is because each harmonic form $\alpha$ can be
uniquely expressed in terms of a holomorphic form $\omega$ as
\be
\alpha = 2\,{\rm Re} \,\omega = \omega + \bar\omega \,.
\ee
Then
\be
*\alpha = -{\rm i} \,\omega + {\rm i} \, \bar\omega \,, 
\ee
and it follows that ${\rm d}\alpha = {\rm d}*\alpha = 0$, since locally 
$\omega=\omega(z){\rm d}z$ for some holomorphic function $\omega(z)$, with
$\bar\omega=\overline{\omega(z)}{\rm d}\bz$ and  $\pr_{\bz}\omega(z) 
= \pr_z\overline{\omega(z)} = 0$.

It is convenient to introduce a canonical basis of the space of 
holomorphic 1-forms $H^0(\Sigma,K_\Sigma)$ \cite{FarKra}.
($K_\Sigma$ 
denotes the canonical sheaf of $\Sigma$.) First, we represent $\Sigma$ as a $4g$-sided 
polygon $\Sigma_{\rm poly}$ with sides identified, as in Fig.~1. (We 
depict the $g=2$ case only, for simplicity.) The 
labelled edges $\{a_j, b_j : 1 \le j \le g\}$ are representatives 
of a canonical (or symplectic) basis of 1-cycles, generating the 
first homology group $H_{1}(\Sigma;\mathbb{Z})\cong\mathbb{Z}^{2g}$.
In this context, the words canonical/symplectic refer to the fact 
that, with respect to this basis, the symplectic 
(i.e.\ skew-symmetric and nondegenerate) pairing 
$\sharp(\cdot,\cdot)$ between homology 1-cycles given
by counting signed intersections with multiplicity has canonical form, namely:
\begin{equation} \label{sympl}
\sharp(a_j,b_k)=\delta_{jk},\quad\sharp(a_j,a_k)=0=\sharp(b_j,b_k), 
\qquad j,k=1,\ldots,g \,.
\end{equation}
Any other canonical basis of $H_{1}(\Sigma;\mathbb{Z})$ is related to 
this one by a linear transformation
in ${\rm Sp}_{2g}(\mathbb{Z})$.
The elements of the canonical basis of holomorphic 1-forms relative 
to this basis of 1-homology are denoted by $\zeta_j, 1 \le j \le g$  
(locally,  $\zeta_j=\zeta_j(z) \, {\rm d}z$, where $\zeta_j(z)$ are 
holomorphic functions of a complex coordinate $z$ in the polygon). 
By definition, they are uniquely determined by the normalisation 
of $a$-periods
\be
\oint_{a_j} \zeta_k = \delta_{jk} \,,
\label{periods1}
\ee
and their $b$-periods are denoted as
\be
\oint_{b_j} \zeta_k = \Pi_{jk} \,.
\label{periods2}
\ee
The $g\times g$ matrix $\Pi$ of $b$-periods has the properties, established by
Riemann, that it is symmetric and its imaginary part ${\rm Im} \, \Pi$ 
is positive definite. Hence ${\rm Im} \, \Pi$ has an inverse 
(which we will use below). Notice that when we change the basis 
of $H_1(\Sigma;\mathbb{Z})$ using some matrix
\be \label{symplblocks}
B=\left( \begin{array}{cc}
B_{11} & B_{12}  \\
B_{21}& B_{22} \end{array} \right)  \ \in \ {\rm Sp}_{2g}\mathbb{Z} \,,
\ee
where $B_{11}, B_{12}, B_{21}$ and $B_{22}$ are $g\times g$ blocks, 
then the matrix of $b$-periods $\Pi$ changes as
\be \label{sympltransf}
\Pi  \mapsto (\Pi\,B_{12}+B_{11})^{-1}(\Pi\,B_{22}+B_{21}) \,,
\ee
where the left factor accounts for a change of basis of 
$H^0(\Sigma,K_\Sigma)$ that is necessary to maintain the 
normalisation of $a$-periods.
\begin{figure}[ht] 
    \begin{center}
	\vspace{.8cm}
	\input{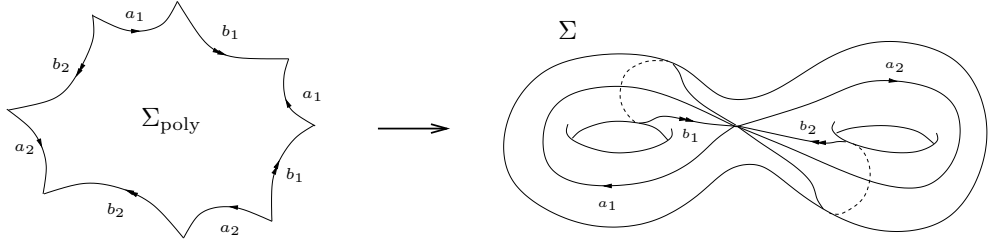}
	\small{
	\caption{Constructing a closed Riemann surface from a polygon}
	}
    \end{center}
\end{figure}\\[-130pt] 
\begin{center}
$\Sigma_{\rm poly}
\qquad\qquad\qquad\qquad\qquad\qquad\qquad\qquad\qquad
\qquad\qquad$\\[-45pt]
$\qquad\qquad\qquad\qquad \Sigma$\\[140pt]
\end{center}

We can now write the time-varying part of the connection, $\alpha$, as
\be
\alpha = \sum_{j=1}^{g}(\overline{c_j} \, \zeta_j 
+ c_j \, \overline{\zeta_j}) \,,
\label{alphaexpan}
\ee
where the coefficients $c_j=c_j(t)$ are complex functions of $t$. Then
\be
e =  \sum_{j=1}^{g}\left(\frac{{\rm d}{\overline{c_j}}}{{\rm d}t} \, \zeta_j 
+  \frac{{\rm d} c_j}{{\rm d}t} \overline{\zeta_j}\right)
\ee
and
\be
*e = \sum_{j=1}^{g}\left(-{\rm i}\frac{{\rm d}{\overline{c_j}}}{{\rm d}t} 
\, \zeta_j + {\rm i}\frac{{\rm d} c_j}{{\rm d}t} \overline{\zeta_j}\right)\,,
\ee
so the kinetic energy of the connection is
\bea
T &=& \half\int_\Sigma e \wedge *e \\
  &=& {\rm i}\sum_{j,k=1}^{g}\frac{{\rm d}{\overline{c_j}}}{{\rm d}t} 
\, \frac{{\rm d} c_k}{{\rm d}t}\int_\Sigma \zeta_j \wedge
  \overline{\zeta_k} \\
  &=& 2\sum_{j,k=1}^{g}({\rm Im} \, \Pi)_{jk} \frac{{\rm d} c_j}{{\rm d}t} \, 
\frac{{\rm d}{\overline{c_k}}}{{\rm d}t} \,, \label{dissolvkin}
\eea
where, to obtain the last line, we have used the result (\cite{FarKra}, p.~65)
\be
\int_\Sigma \zeta_j \wedge \overline{\zeta_k} = -2{\rm i}({\rm Im} \, \Pi)_{jk} \,,
\label{intzetbarzet}
\ee
and the symmetry of ${\rm Im} \, \Pi$.

The expression (\ref{dissolvkin}) can be interpreted as the
kinetic energy of dissolved vortices, and arises purely from the
electric field of the time-varying connection. It only depends on 
the complex structure of $\Sigma$. Exactly where any
vortices are, and how they are moving, will be clarified below.
The metric on the moduli space of dissolved vortices is
obtained by dropping a factor $\half$, and is therefore \cite{Nas} 
\be
{\rm d}s^2 = 4 \sum_{j,k=1}^{g}({\rm Im} \, \Pi)_{jk} \, 
{\rm d}c_j \, {\rm d}{\overline{c_k}} \,,
\label{flatmetric}
\ee
a flat metric on a space of complex dimension $g$. 

The final point is that the range of the coordinates $c_j$ is not the 
whole of $\C^g$: two connections with the same curvature are (globally) gauge
equivalent if and only if they
have the same holonomies around all 1-cycles. The connections ${\rm d}_a$ and
${\rm d}_{a+\alpha}$, where $\alpha$ is a harmonic 1-form, have the same 
holonomies if the integral of $\alpha$ around any 1-cycle on $\Sigma$ is 
an integer multiple of $2\pi$, that is, if $d_{\alpha}$ has trivial holonomy. 
This condition is equivalent to requiring the integrals around the
cycles $a_j$ and $b_j$ to be integral multiples of $2\pi$. For
$\alpha =\sum_{j=1}^{g} (\overline{c_j} \, \zeta_j 
+ c_j \, \overline{\zeta_j})$, these integrals are
\bea
\oint_{a_j} \alpha &=& \overline{c_j} + c_j \,, \\
\oint_{b_j} \alpha &=& \sum_{k=1}^{g}(\Pi_{jk} \, \overline{c_k} + 
\overline{\Pi_{jk}} \, c_k)  \,.
\eea
The 1-forms $\alpha$ representing connections with trivial holonomy therefore
form a lattice of rank $2g$ inside the space of global harmonic 
1-forms ${\mathcal H}_1(\Sigma;\mathbb{R})=\mathbb{R}^{2g} \cong \C^g$, 
defined by the $2g$ real conditions, 
\be
\frac{1}{\pi}{\rm Re} \, c_j \in \Z \,, \quad 
\frac{1}{\pi}{\rm Re} \,\sum_{k=1}^{g} (\Pi_{jk} \, \overline{c_k}) \in \Z \,.
\label{cLattice}
\ee
So the moduli space of connections describing dissolved vortices is $\C^g$ 
modulo this lattice, which is a complex torus~\cite{GriHar} with the flat 
metric (\ref{flatmetric}). Its volume is $(2\pi)^{2g}$ \cite{Nas}.

This torus is essentially the Jacobian of $\Sigma$, as the lattice here 
is related to the period lattice in $\C^g$ generated by 
the column vectors of the unit matrix and the column vectors of the 
matrix $\Pi$, as we shall show next. 
More precisely, this torus is dual to the Jacobian, but will be naturally identified with the
Jacobian once a normalisation constant is chosen.
Notice that all these results, 
and the metric (\ref{flatmetric}) in particular, are independent of the
vortex number $N$ (which only appeared via the reference 
connection ${\rm d}_a$).

\section{Geometry of the Jacobian variety} \label{geomJac}
\news

Recall that the vector space $H^0(\Sigma,K_\Sigma)$ of holomorphic 1-forms on
$\Sigma$ is of complex dimension $g$. 
One can embed the group $H_1(\Sigma;\mathbb{Z})$ of homology 1-cycles into 
the dual space $H^0(\Sigma,K_\Sigma)^*$
as follows: for each 1-cycle  $\gamma$, we consider the
linear functional $\{\gamma\} \in H^0(\Sigma,K_\Sigma)^*$ defined by
\be
\omega \mapsto \oint_{\gamma} \omega \,,
\ee
sending a holomorphic 1-form $\omega$ to its line integral around 
$\gamma$. Let $\Lambda$ denote the image of this embedding, which is 
a lattice of rank $2g$ in $H^0(\Sigma,K_\Sigma)^*$
generated by the elements $\{a_j\}$, $\{b_j\}$,
\be
\omega \mapsto \oint_{a_j} \omega \,, \quad  \omega \mapsto \oint_{b_j} \omega
\ee
with $\{a_j, b_j : 1 \le j \le g\}$ the basis of homology 1-cycles introduced
earlier. By definition, the Jacobian of $\Sigma$ is the quotient
\be
{\rm Jac}(\Sigma)=H^0(\Sigma,K_\Sigma)^*/\Lambda \,, \qquad \Lambda = \{ H_1(\Sigma;\mathbb{Z}) \}
\ee
and is a complex torus of real dimension $2g$. This manifold comes equipped 
with a complex structure, induced from multiplication by ${\rm i}$ 
in $H^0(\Sigma,K_\Sigma)$, which in turn comes from the complex
structure $j_\Sigma$ on the Riemann surface $\Sigma$. Clearly, ${\rm Jac}(\Sigma)$ is also 
an analytic abelian Lie group,
with operation induced from the addition in the vector space $H^0(\Sigma,K_\Sigma)^*$.

Once we fix the canonical basis of holomorphic 1-forms 
$\{\zeta_k : 1 \le k \le g\}$, the vector space
$H^0(\Sigma,K_\Sigma)^*$ acquires natural complex coordinates 
$\chi_k :H^0(\Sigma,K_\Sigma)^*\rightarrow \mathbb{C}$ (for $1 \le k \le g$), 
which are defined  by extending linearly (over $\mathbb{R}$) the functionals
\be
\{\gamma\}  \mapsto    \oint_{\gamma} \zeta_k =: \chi_k(\{ \gamma \}) \,,
\ee
defined on the image $\Lambda\cong\mathbb{Z}^{2g}$ 
of $ H_1(\Sigma;\mathbb{Z})$, to the whole of
$H^0(\Sigma,K_\Sigma)^*\cong \Lambda\otimes_{\mathbb{Z}}\mathbb{R}$;
in particular,
\bea
\chi_k(\{a_j\})&=& \oint_{a_j} \zeta_k = \delta_{jk} \,, \\
\chi_k(\{b_j\})&=& \oint_{b_j} \zeta_k = \Pi_{jk} \,.
\eea
In these coordinates, the lattice $\Lambda$ is generated over 
$\mathbb{Z}$ by the $2g$
vectors in $\C^g$ which are the column vectors of the unit matrix and
the column vectors of the matrix $\Pi$, and ${\rm Jac}(\Sigma)$ has the 
explicit form $\C^g/\Lambda$.

The projectively flat connections ${\rm d}_{a + \alpha}$ that occur in 
the limit of dissolved vortices, discussed in the previous section, 
also define elements of $H^0(\Sigma,K_\Sigma)^*$, via wedge product and integration over
$\Sigma$. (This is a variant of a construction usually applied to flat
connections.) Specifically, once we fix a projectively flat 
connection ${\rm d}_a$, we can identify another projectively 
flat connection ${\rm d}_{a + \alpha}$ with the harmonic 1-form 
$\alpha$, which in turn defines an element
of $H^0(\Sigma,K_\Sigma)^*$ via
\be
\omega \mapsto \frac{1}{2\pi} \int_\Sigma \alpha \wedge \omega \,,
\label{connJ}
\ee 
whose coordinates are
\be
\chi_k = 
\frac{1}{2\pi} \int_\Sigma \alpha \wedge \zeta_k \,.
\ee
In (\ref{connJ}) we are introducing a normalisation factor of $2 \pi$ for convenience.
Using the expansion (\ref{alphaexpan}), and the integral 
(\ref{intzetbarzet}), and noting that $\zeta_j \wedge \zeta_k$ 
vanishes, we deduce that
\be
\chi_k = \frac{\rm i}{\pi}\sum_{l=1}^{g}({\rm Im} \, \Pi)_{kl} c_l \,.
\label{chi-to-c}
\ee
Since ${\rm Im} \, \Pi$ is invertible, the coordinates $\chi_k$ or
$c_k$ are equally good at parametrising the element of $H^0(\Sigma,K_\Sigma)^*$
corresponding to $\alpha$.

Now recall that the coordinates $c_k$ are defined only modulo the 
lattice specified by the conditions (\ref{cLattice}). These conditions 
are equivalent to
\be
\overline{c_j} + c_j = 2\pi m_j \,, 
\quad \sum_{k=1}^{g}(\Pi_{jk} \, \overline{c_k} +
\overline{\Pi_{jk}} \, c_k )= 2\pi n_j \,,
\ee
where $\{m_j,n_j: 1 \le j \le g\}$ are integers. Eliminating
$\overline{c_k}$, they become
\be
\frac{\rm i}{\pi}\sum_{k=1}^{g}({\rm Im} \, \Pi)_{jk} c_k 
= \sum_{k=1}^{g}(\Pi_{jk}m_k - \delta_{jk}n_k) \,,
\ee
and in terms of the coordinates on $H^0(\Sigma,K_\Sigma)^*$ defined by
(\ref{chi-to-c}) they take the form
\be
\chi_j = \sum_{k=1}^{g}(\Pi_{jk}m_k - \delta_{jk}n_k) \,,
\ee
a vector of coordinates that is an integer
combination of the column vectors of the unit matrix and the column
vectors of the matrix $\Pi$. But this is precisely a vector in the lattice 
$\Lambda$. So, for a connection ${\rm d}_{a + \alpha}$, 
equation (\ref{connJ}) defines 
unambiguously an element in the quotient space $H^0(\Sigma,K_\Sigma)^*/\Lambda$, 
which is the Jacobian. 

We had already pointed out that the moduli space 
of projectively flat connections is a (real) $2g$-torus, hence diffeomorphic 
to ${\rm Jac}(\Sigma)$. We can interpret equation (\ref{chi-to-c}) as
giving an explicit 
diffeomorphism between the two tori (the Jacobian and its dual) expressed in local coordinates 
and use it to pull back the Riemannian metric (\ref{flatmetric}) on 
the moduli space of projectively flat connections to obtain a 
metric $G_{\rm J}$ on ${\rm Jac}(\Sigma)$: in the coordinates above,
\be
G_{\rm J} = 4\pi^2\sum_{j,k=1}^{g}({\rm Im} \, \Pi)^{-1}_{jk} \, 
{\rm d}\chi_j {\rm d}\overline{\chi_k} \,.
\label{metricJ}
\ee
It is important to note that this metric is intrinsic to 
${\rm Jac}(\Sigma)$; since it is invariant under translations, it does 
not depend on the choice of reference projectively flat 
connection ${\rm d}_a$. Now there is another intrinsic geometric 
structure on ${\rm Jac}(\Sigma)$, namely, a symplectic
form that descends from the translation-invariant symplectic 
form on $H^0(\Sigma,K_\Sigma)^*$ defined by linearly extending (over
$\mathbb{R}$) the intersection pairing (\ref{sympl}). 
We denote it by $\Omega_{\rm J}(\cdot,\cdot)$. From $G_{\rm J}$ 
and $\Omega_{\rm J}$, we can define a hermitian metric 
$H_{\rm J}$ on ${\rm Jac}(\Sigma)$ by setting
\be \label{hermpol}
H_{\rm J}(\cdot,\cdot)=G_{\rm J}(\cdot,\cdot)
+{\rm i}\Omega_{\rm J}(\cdot,\cdot) \,.
\ee
Since $\Omega_{\rm J}$ is closed, there is an underlying K\"ahler 
structure, and one can recover any of $G_{\rm J}$, $\Omega_{\rm J}$ 
and $H_{\rm J}$ from just one of these structures and the complex structure
(multiplication by ${\rm i}$). Another way to see that this K\"ahler 
structure is intrinsic is to note that its pullback to 
$H^0(\Sigma,K_\Sigma)^* \cong\mathbb{C}^g$ coincides with the positive definite 
hermitian form induced from the nondegenerate pairing on $H^0(\Sigma,K_\Sigma)$ given by
\be \label{1formsL2}
(\zeta,\omega) \mapsto  \int_\Sigma \zeta \wedge \bar{\omega} \,.
\ee
This bilinear form on $H^0(\Sigma,K_\Sigma)^*$ has the property of being 
$\mathbb{Z}$-valued on elements of the lattice $\Lambda$ (this 
follows from (\ref{sympl}) and Poincar\'e duality),
and is called a polarisation of the torus 
${\rm Jac}(\Sigma)=H^0(\Sigma,K_\Sigma)^*/\Lambda$. Since the matrix obtained by
evaluating the imaginary part on a basis of $\Lambda$ has unit 
determinant, one speaks of a principal polarisation. Notice that 
polarisations are defined in terms of the matrix of $b$-periods
$\Pi$ only up to transformations of the form (\ref{sympltransf}) 
in terms of the blocks in (\ref{symplblocks}), that is, they are 
parametrised by the orbits of ${\rm Sp}_{2g}(\mathbb{Z})$ acting on 
the space of symmetric, complex $g\times g$ matrices with positive 
definite imaginary part. Such a transformation relates hermitian forms 
of the form (\ref{hermpol}) that are isometric (i.e. pull-backs of 
each other under linear transformations).

The Jacobian plays another important role, that of classifying
holomorphic line bundles over $\Sigma$. This classification is finer than that
of (smooth) line bundles without a holomorphic structure specified.  
For fixed $N$, all line bundles are topologically equivalent, 
but, as we now recall \cite{GriHar,Jost}, the moduli space of 
holomorphic line bundles on $\Sigma$ is a copy of ${\rm Jac}(\Sigma)$. 

A holomorphic line bundle ${\mathcal L} \rightarrow \Sigma$ of first Chern 
class (or degree) $N$ is determined by a divisor class of degree $N$. 
Such a class is represented by divisors on $\Sigma$, consisting of formal 
sums of points of $\Sigma$ written as 
\be\label{divisor}
\sum_{j=1}^{n+N}P_j  - \sum_{k=1}^{n}Q_k
\ee
for some $n\in \mathbb{N}_0$. The points $P_j$ with positive
coefficients are locations of zeros of a meromorphic section of the 
bundle, whereas the points with negative coefficients $Q_k$ are
locations of the poles of the same section. (Note
that the points appearing in a divisor need not be distinct, to
account for poles or zeros of higher multiplicity; if they are
distinct, the divisor is said to be reduced.)
Two divisors of degree $N$ belong to the same (linear) class, 
i.e.\ represent the same holomorphic line bundle of degree $N$, 
if their difference is a divisor of degree zero describing 
the locus of zeros and poles of some meromorphic function on $\Sigma$.

The relation of divisors on $\Sigma$ with the Jacobian is achieved by 
the Abel--Jacobi map. To define this, one needs to choose a base 
point $P_0$ on $\Sigma$. Then the Abel--Jacobi image of a divisor 
(\ref{divisor}) representing ${\mathcal L}$ is the element 
of ${\rm Jac}(\Sigma)$ defined through the divisor's action on 
holomorphic 1-forms via abelian integrals,
\be
\omega \mapsto \sum_{j=1}^{n+N} \int_{P_0}^{P_j} \omega -    
\sum_{k=1}^n \int_{P_0}^{Q_k} \omega \,.
\ee
Equivalently, its coordinates on the Jacobian are
\be
\chi_l = \sum_{j=1}^{n+N} \int_{P_0}^{P_j} \zeta_l -    
\sum_{k=1}^n \int_{P_0}^{Q_k} \zeta_l \,.
\ee
These are well defined in ${\rm Jac}(\Sigma)$ despite the non-uniqueness 
of the paths from $P_0$ to $P_i$ and $Q_i$, since the ambiguities 
are contained in $\Lambda$. Also, divisors which map to the same point 
in the Jacobian are in one divisor class, as by
Abel's theorem they correspond to the same line bundle. The 
Jacobi inversion theorem implies that to each point in the Jacobian 
there is a non-trivial divisor class, and hence a holomorphic line 
bundle. Notice that choosing a different base point $P_0 \in \Sigma$ 
amounts to a global translation in ${\rm Jac}(\Sigma)$. In summary, 
the moduli space of holomorphic line bundles on $\Sigma$ of degree $N$ 
(i.e. with fixed but arbitrary vortex number $N$) is a copy of the 
Jacobian of $\Sigma$.

In the dissolving limit, described at the end of
Section~\ref{dynamics}, vortices are obtained from line bundles of degree 
$N$ with holomorphic sections, and these are not generic. They are the 
bundles that arise from effective divisors, i.e., those with a 
divisor of zeros $P_1 + P_2 +\dots + P_{N}$ but vanishing divisor of 
poles. The coordinates on the Jacobian corresponding such a divisor are
\be
\chi_l= \sum_{j=1}^{N} \int_{P_0}^{P_j} \zeta_l \,.
\label{effdivJac}
\ee  
The space of effective divisors of degree $N$ is clearly 
${\rm Sym}^N \Sigma$. If $N<g$, its image in ${\rm Jac}(\Sigma)$ is denoted 
$W_N$, which is a complex subvariety of ${\rm Jac}(\Sigma)$ of codimension $g-N$, 
whereas for $N\ge g$ it is the whole of the Jacobian.
Of most interest to us is the set of effective divisors of
degree 1, representable by  single points $ P $. This is a copy of $\Sigma$
itself, and the Abel--Jacobi map embeds $\Sigma={\rm Sym}^1(\Sigma)$ in 
${\rm Jac}(\Sigma)$ as a smooth complex curve $W_1$. The coordinates of 
the image in ${\rm Jac}(\Sigma)$ of the degree one divisor $P $  are
\be
\chi_j = \int_{P_0}^{P} \zeta_j \,.
\label{effdivPJac}
\ee
In the next section we show that the Abel--Jacobi image of $\Sigma$ 
captures the geometry of the one-vortex moduli space in the 
dissolving limit. The generic case when $1<N<g$ is 
that $W_N$ has singularities at points of ${\rm Jac}(\Sigma)$ 
(i.e. line bundles) that admit linearly independent holomorphic sections.
We shall discuss their significance further in 
Sections~\ref{mvortices} and~\ref{singular}.

\section{One vortex in the dissolving limit}  \label{1vortex}
\news

In the regime that we call the limit of dissolving vortices, which 
models the situation where $\tau A - 4\pi N$ is small and positive, 
we shall adopt the following prescription to approximate solutions 
to the vortex equations. From the integrated second vortex equation 
(\ref{Bogo2int}) we see that $\phi$ is small in the $L^2$-norm, and 
we therefore make the approximation that the magnetic field is the 
same as for the situation of dissolved vortices discussed in 
Section~\ref{dissolved}. However, the need to solve 
the first vortex equation puts constraints on the connection. To 
investigate this it is helpful to transform from a unitary to a 
holomorphic gauge, where holomorphic sections of 
${\mathcal L}\rightarrow \Sigma$ are locally given by holomorphic functions.

To make possible a transition to holomorphic gauge, one extends the
class of gauge transformations, so that they are valued in the 
complexification ${\rm U}(1)^{c}=\C^*$ rather than ${\rm U}(1)$. 
Complex gauge transformations are still given by the formulae 
(\ref{gaugetrm1}) and (\ref{gaugetrm2}), but $\chi$ can now take 
complex values. The first vortex equation is invariant under 
complexified gauge transformations, but the second 
vortex equation generally is not, since it involves the
hermitian structure of ${\mathcal L}\rightarrow \Sigma$, which should be 
multiplied by the quantity $|{\rm e}^{{\rm i}\chi}|^2$ under
a complex gauge transformation ${\rm e}^{{\rm i}\chi}: 
\Sigma\rightarrow \mathbb{C}^*$, so that covariance is preserved.
However, we are replacing the second vortex equation by the 
equation (\ref{projflatf}) of projectively
flat connections, and this is invariant under the complexified gauge group.

Let us start with a $\mathbb{C}^*$-connection ${\rm d}_{a}$, locally 
given by a connection 1-form $a=a_z {\rm d}z + a_{\bz}{\rm d}\bz$ 
for which the reality condition $a_{\bar{z}} = \overline{a_{z}}$ 
need not hold. Using a complex gauge transformation, it is always
possible to go to a holomorphic gauge, where $a_{\bz} = 0$ 
(in each trivialising chart). In this gauge, the transition functions 
for the line bundle ${\mathcal L}\rightarrow \Sigma$ are holomorphic functions
from overlaps of trivialising patches to $\mathbb{C}^*$. 
For the connection 1-form (\ref{connK}), we can be more explicit. After making
the gauge transformation 
\be
\chi = -\frac{{\rm i} \tau}{4} \K \,,
\label{unitohol0}
\ee
we obtain
\be
a = -\frac{{\rm i} \tau}{2} \, \pr \K  
= -\frac{{\rm i} \tau}{2} \, \pr_z \K \, {\rm d}z \,,
\label{connholK}
\ee
so that $a_{\bz}=0$.

In holomorphic gauge, the first vortex equation reduces to
\be
\bar\pr \phi = 0 \,,
\ee
as the holomorphic structure operator $\bar\pr_a$ coincides with the
$\bar\pr$-operator defined from the complex structure of $\Sigma$ alone;
that is, on a trivialisation $\phi$ is a holomorphic function. 
If ${\mathcal L}$ has no
nontrivial holomorphic section then $\phi$ must vanish everywhere, and
there is no vortex, so we exclude such bundles from further consideration.
For ${\mathcal L}$ to have a holomorphic section, its divisor class 
must contain an effective divisor, $P_1+ P_2 + \dots +P_{N}$
say. Then $\phi$ vanishes at these isolated points (with higher
multiplicity, whenever some of the points $P_j$ coincide), and they 
are the vortex centres. 

Let us now focus on the one-vortex case, with $N=1$. The divisor class
of ${\mathcal L}$ must be represented by an effective divisor $P$, 
which is just any point of $\Sigma$, and $\phi$ has a simple zero at
$P$. The moduli space $\M_1$ of one-vortex solutions is therefore $\Sigma$.

Our present task is to understand these bundles more concretely, so as
to calculate the metric on $\M_1$. To do this, it is
convenient to first fix the holomorphic line bundle 
${\mathcal L}={\mathcal L}_0$, with divisor $P_0$, associated with 
having one vortex with centre at the base point $P_0 \in \Sigma$. We fix 
the connection 1-form $a$ on a trivialising patch of
${\mathcal L}_0$ to be of the form (\ref{connholK}). 
Next we consider moving the vortex to the point 
$P$. This will require changing the holomorphic bundle 
to ${\mathcal L}={\mathcal L}_P$, associated with the divisor $P$ 
and different transition functions. Alternatively, it will correspond to 
keeping the bundle and transition functions fixed but replacing the 
connection by ${\rm d}_a-{\rm i}\alpha$, with $\alpha$ a particular harmonic 
1-form that depends on $P$ and $P_0$. By relating these two points of 
view, we  will be able to calculate how $\alpha$ depends on $P$, and from this 
find the kinetic energy of the moving vortex.   

For the first point of view, we assume that $\Sigma$ is represented by the 
polygon $\Sigma_{\rm poly}$ in Fig. 2 (with $P_0$ and $P$ not on the
edges), and that our line bundles are trivialised over this (open)
polygon. Introduce a complex coordinate $z:\Sigma_{\rm poly} \rightarrow \mathbb{C}$
(with $z=0$ at some interior point distinct from $P_0$ and $P$), and let
$Z_0=z(P_0)$ and $Z=z(P)$. ${\mathcal L}_0$ is defined by certain 
(holomorphic) transition functions connecting (neighbourhoods of) the 
paired edges $a_j, a^{-1}_j$ and $b_j, b^{-1}_j$. To obtain the bundle 
${\mathcal L}_P$ it is sufficient to change the transition functions
by constant factors on each edge. This is a choice of gauge, and is 
satisfactory because we only want to change the holonomy of the 
connection, without changing the
magnetic field. Let us call these factors ${\rm e}^{\mu_j}$ (the additional
factor connecting $a_j$ to $a^{-1}_j$) and ${\rm e}^{\nu_j}$ (the additional
factor connecting $b_j$ to $b^{-1}_j$).
\begin{figure}[ht] 
    \begin{center}
	\vspace{.8cm}
	\input{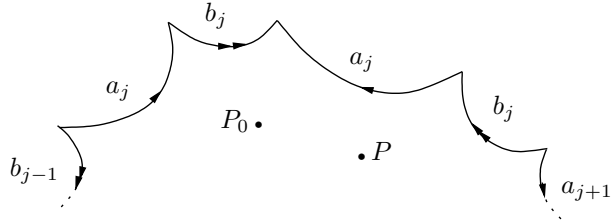}
	\small{
	\caption{The polygonal trivialising patch $\Sigma_{\rm poly}$}
	}
    \end{center}
\end{figure}\\[-105pt]
\begin{center}
$P_0 \,${\tiny $\bullet$}$ \qquad\qquad$ \\
$\qquad\qquad\qquad ${\tiny$\bullet$}$\,  P$\\[70pt]
\end{center} 

The dependence of these factors on $P_0$ and $P$ can be determined 
using abelian differentials. Let $\xi^{P_0}_{P}$ be an abelian 
differential of the third kind (a meromorphic 1-form on $\Sigma$ 
with only simple poles \cite{FarKra}), whose poles are located 
at $P_0$ and $P$ with residues $-1$ and $1$, respectively. Such 
a differential exists 
and is unique up to addition of a holomorphic 1-form (an abelian 
differential of the first kind). Define the locally holomorphic function
\be
f(z) = - \int_0^z \xi^{P_0}_P \,.
\ee
$f$ has singularities at $P_0$ and $P$ and its value also depends on the
homology class of the integration path in $\Sigma_{\rm poly} 
\setminus  \{P_0 , P\}$. The singularities are of type 
$-\log(z - Z_0)$ and $\log(z - Z)$, so $f$ branches at these points and
is multivalued on $\Sigma_{\rm poly} \setminus  \{P_0 , P\}$, the 
differences of values at each point being integer
multiples of $2 \pi {\rm i}$. If we now set 
${H}^{P_0}_{P}(z) := {\rm e}^{f(z)}$, we obtain a single-valued 
holomorphic function on $\Sigma_{\rm poly}$ which has a simple pole at 
$P_0$ and a simple zero at $P$. Let $\phi_0$ be the holomorphic 
section of ${\mathcal L}_0$  with its zero at $P_0$. Then 
$H^{P_0}_P \phi_0|_{\Sigma_{\rm poly}}$ has a zero 
at $P$ but no longer has a zero at $P_0$. So $H^{P_0}_{P} \phi_0$ 
is a good candidate for the holomorphic section of ${\mathcal L}_P$. 

At this point, we note that $\phi_0$ and $H^{P_0}_P \phi_0$ are 
indeed sections of different line bundles, because $H^{P_0}_P$ does 
not extend to a meromorphic function on $\Sigma$ when the edges 
of $\Sigma_{\rm poly}$ are identified (if it did extend
to a meromorphic function with a single zero and a single pole, 
$\Sigma$ would necessarily be rational with $g=0$). By comparing $H^{P_0}_{P}$ 
at equivalent points of the paired edges $\{a_j, a^{-1}_j\}$ 
and $\{b_j, b^{-1}_j\}$, we see that the action of $H^{P_0}_P$ 
on $\phi_0$ amounts to including additional transition factors 
${\rm e}^{\mu_j}$ and ${\rm e}^{\nu_j}$ in the transition 
functions of the bundle ${\mathcal L}_0 \rightarrow \Sigma$, namely
\be \label{transitfact}
e^{\mu_j} = e^{\oint_{b_j} \xi^{P_0}_P} \,, \quad
e^{\nu_j} = e^{- \oint_{a_j} \xi^{P_0}_P} \,,
\ee
because a path in $\Sigma_{\rm poly}$ connecting equivalent points on edges $\{a_j,
a^{-1}_j\}$ is homologous to the edge $b_j$, and similarly a path 
connecting equivalent points on edges $\{b_j,b^{-1}_j\}$ is
homologous to the edge $a^{-1}_j$. These factors are constant, because
they are independent of the choice of equivalent points on the paired 
edges, by Cauchy's theorem. They also have no dependence on the
homology classes of the paths 
of integration, provided the paths avoid $P_0$ and $P$, since the
residues at the poles of $\xi^{P_0}_P$ are $\pm 1$.
So we see that multiplication by this global object $H^{P_0}_P$ 
transforms holomorphic sections of ${\mathcal L}_0 \rightarrow \Sigma$ 
into holomorphic sections of ${\mathcal L}_P \rightarrow \Sigma$. In the language
of the theory of holomorphic vector bundles over curves, $H^{P_0}_P$
describes the action (on holomorphic sections) of a double Hecke 
modification~\cite{FreBen} that shifts the twisting of a holomorphic line bundle 
of degree one from $P_0$ to $P$.

Now recall  that $\xi^{P_0}_P$ is not unique, because one may add to it any
holomorphic 1-form without changing its poles and residues. Let us
choose $\xi^{P_0}_P$ to be normalised by the condition that its integrals 
along all the edges $a_j$ of $\Sigma_{\rm poly}$ vanish. In bundle terms, 
this corresponds to fixing some of the remaining freedom to perform 
holomorphic gauge transformations. For this normalised $\xi^{P_0}_P$,
\be
e^{\mu_j} = e^{\oint_{b_j} \xi^{P_0}_P}
\ee
and $\nu_j\equiv 0 \,({\rm mod} \,2 \pi {\rm i})$. A Riemann bilinear 
relation gives \cite{FarKra}
\be
\oint_{b_j} \xi^{P_0}_P = -2\pi{\rm i} \int_{P_0}^P \zeta_j \,,
\ee
where $\zeta_j$ is the $j$th canonical holomorphic 1-form.
(This is derived by applying the residue theorem to a product of 
$\xi^{P_0}_P$ and the indefinite integral of $\zeta_j$, integrated
around the boundary of $\Sigma_{\rm poly}$.) Therefore
\be
{\rm e}^{\mu_j} = {\rm e}^{-2\pi{\rm  i} \int_{P_0}^P \zeta_j} \,,
\ee
or, as an equation relating the exponents,
\be
\mu_j = -2\pi{\rm  i} \int_{P_0}^P \zeta_j 
\quad ({\rm mod} \; 2 \pi {\rm i}) \,.
\label{formulamuj}
\ee

Recalling (\ref{effdivPJac}), we see that, rather remarkably, 
$\frac{\rm i}{2\pi} \mu_j$ are the coordinates 
$\chi_j$ of the image of $P$ in ${\rm Jac}(\Sigma)$ by the Abel--Jacobi 
map with base-point $P_0$. So, the holomorphic line bundle 
${\mathcal L}_P$ with effective divisor $P$ is obtained from the
holomorphic line bundle ${\mathcal L}_0$ with effective divisor $P_0$ using a 
change in transition functions which is directly related to the 
Abel--Jacobi image of $P$. (This gives yet another
characterisation of the Jacobian, as the moduli space of bundle
transition functions --- the so-called theta characteristics.)

For the second point of view, we return to the bundle ${\mathcal L}_0$
(trivialised over $\Sigma_{\rm poly}$), with connection 1-form $a$ and 
holomorphic section $\phi_0$ describing a vortex centred at $P_0$, 
and now look for the connection 1-form $a+\alpha$, where $\alpha$ is 
the restriction of a global, real harmonic 1-form, that gives 
a vortex centred at $P$. We first transform to the gauge discussed 
above, where 
$a = -\frac{i\tau}{2} \pr_z \K \, {\rm d}z$. In this
gauge, the first vortex equation in our trivialisation is
\be
\pr_{\bz} \phi -{\rm  i} \alpha_{\bz} \phi = 0 \,,
\ee
so the function $\Sigma_{\rm poly} \rightarrow \mathbb{C}$ representing 
the section $\phi$ is not holomorphic. Now recall that $\alpha = 
\omega +\overline{\omega}$, where $\omega = \omega(z) \, {\rm d}z$ 
extends to a global holomorphic 1-form. Let
\be
\chi(\bz) = -\overline{\int_0^z \omega} \,,
\label{unitohol1}
\ee
which is an anti-holomorphic function on $\Sigma_{\rm poly}$. The complex 
gauge transformation $e^{{\rm i}\chi}$ makes $\alpha_{\bz}$ vanish on
$\Sigma_{\rm poly}$. Its effect is therefore to transform the
complete connection ${\rm d}_a - {\rm i}\alpha$ to holomorphic gauge, 
but at the expense  
of changing the bundle transition functions, as $e^{{\rm i}\chi}$ is not
single-valued on $\Sigma$. The additional factors depend on
the difference between the values of $\chi$ at equivalent points 
on the paired edges. Before writing these down, we need to allow 
for the possibility of a further holomorphic gauge transformation, 
leaving us in holomorphic gauge. We therefore modify
(\ref{unitohol1}) to
\be
\chi(z,\bz) = -\overline{\int_0^z \omega} +
\int_0^z \theta
\ee
where $\theta$ extends to a global holomorphic 1-form. The additional factors
defining the new bundle transition functions are $e^{\mu_j}$ and
$e^{\nu_j}$, where
\bea
\mu_j &=&{\rm  i}\overline{\oint_{b_j} \omega} -{\rm i}\oint_{b_j} \theta \,,
\\
\nu_j &=& -{\rm i}\overline{\oint_{a_j} \omega} + {\rm i}\oint_{a_j} \theta \,.
\eea
As before, these factors are constants, by Cauchy's theorem.

Now expand the 1-forms $\omega$ and $\theta$
in terms of the canonical basis of holomorphic 1-forms as
\be
\omega = \sum_{k=1}^{g}\overline{c_k} \, \zeta_k \,, 
\quad \theta = \sum_{k=1}^{g}d_k \,
\zeta_k \,.
\ee
Then, in terms of the periods (\ref{periods1}) and (\ref{periods2}),
\bea
\mu_j &=& {\rm i} \sum_{k=1}^{g}\overline{\Pi_{jk}} \, 
c_k - {\rm i}\sum_{k=1}^{g}\Pi_{jk} d_k \,,
\\
\nu_j &=& -{\rm i}\,c_j + {\rm i}\, d_j \,.
\eea
Choose $\theta$ so that $\nu_j$ vanishes, i.e. $d_j = c_j$. Then
\be
\mu_j = {\rm  i} \sum_{k=1}^{g}(\overline{\Pi_{jk}} - \Pi_{jk}) c_k 
      = 2\sum_{k=1}^{g}({\rm Im} \, \Pi)_{jk} \, c_k \,.
\ee
Since the matrix ${\rm Im} \, \Pi$ is invertible,
\be 
c_k = \half\sum_{j=1}^{g} ({\rm Im} \, \Pi)_{jk}^{-1} \mu_j \,.
\ee
Now we can use our earlier result (\ref{formulamuj}), determining 
the factors $\mu_j$ which produce a holomorphic section vanishing at 
$P$. This leads to our final expression for the coefficients,
\be
c_k = -\pi {\rm i}\sum_{j=1}^{g} ({\rm Im} \, \Pi)_{jk}^{-1} 
\int_{P_0}^P \zeta_j \,.
\label{c_kP}
\ee
The harmonic 1-form $\alpha$ that shifts a vortex from $P_0$ to $P$,
in the unitary gauge, $\alpha = \sum_{k=1}^{g}(\overline{c_k} \, \zeta_k 
+ c_k \, \overline{\zeta_k})$, is therefore
\be
\alpha = \pi {\rm i}\sum_{j,k=1}^{g}({\rm Im} \, \Pi)_{jk}^{-1}
\left[\left(\overline{\int_{P_0}^P \zeta_j}\right) \zeta_k 
- \left(\int_{P_0}^P \zeta_j\right)\overline{\zeta_k}\right] \,.
\label{connP_OPshift}
\ee

This harmonic 1-form is exactly what one would expect from the
discussion of the Jacobian in the previous section. The degree one divisor
$P$ corresponds to the point on ${\rm Jac}(\Sigma)$ with coordinates $\chi_j =
\int_{P_0}^{P} \zeta_j$, by (\ref{effdivPJac}). The related 
coordinates $c_j$ are therefore
\be
c_j = -\pi{\rm i}\sum_{k=1}^{g}({\rm Im} \, \Pi)_{jk}^{-1}
\int_{P_0}^{P} \zeta_k \,,
\label{c_jJac}
\ee
by the inverse of (\ref{chi-to-c}). This agrees with (\ref{c_kP}).

Using the expression for $\alpha$, we can find the electric field 
associated with a moving vortex. Let us first note the infinitesimal change in
$\alpha$ that occurs when the vortex is shifted from $P$ to a
neighbouring point which we call $P + \delta P$. Let the coordinates 
of $P$ and $P + \delta P$ be $Z$ and $Z + \delta Z$. Then the change 
in $\alpha$ is
\be
\delta \alpha = \pi{\rm  i} \sum_{j,k=1}^{g}({\rm Im} \, \Pi)_{jk}^{-1}
\left[\overline{\zeta_j(Z)} \, \delta\bZ \, \zeta_k -
  \zeta_j(Z) \,\delta Z \, \overline{\zeta_k}\right] \,.
\ee
Therefore, the 1-form electric field of a moving vortex is
\be
e = \pi{\rm i}\sum_{j,k=1}^{g} ({\rm Im} \, \Pi)_{jk}^{-1}
\left[\overline{\zeta_j(Z)}\frac{{\rm d}\bZ}{{\rm d}t} \zeta_k 
- \zeta_j(Z)\frac{{\rm d}Z}{{\rm d}t} \overline{\zeta_k}\right] \,,
\ee
with Hodge dual
\be
*e = \pi \sum_{j,k=1}^{g} ({\rm Im} \, \Pi)_{jk}^{-1}
\left[\overline{\zeta_j(Z)}\frac{{\rm d}\bZ}{{\rm d}t} \zeta_k 
+ \zeta_j(Z)\frac{{\rm d}Z}{{\rm d}t} \overline{\zeta_k}\right] \,,
\ee
so the kinetic energy of the moving vortex is
\bea
T &=& \half \int_\Sigma e \wedge *e \\
  &=& \half \pi^2 {\rm i} \sum_{j,k,m,n=1}^{g} 
({\rm Im} \, \Pi)_{jk}^{-1}({\rm Im} \, \Pi)_{mn}^{-1}
      \bigg[ \overline{\zeta_j(Z)}\zeta_m(Z) \int_\Sigma \zeta_k \wedge 
      \overline{\zeta_n} \nonumber \\
  &&  \quad\quad\quad\quad\quad\quad\quad\quad 
      - \, \zeta_j(Z)\overline{\zeta_m(Z)} 
      \int_\Sigma \overline{\zeta_k} \wedge \zeta_n \bigg] \,
      \frac{{\rm d}Z}{{\rm d}t}\frac{{\rm d}\bZ}{{\rm d}t} \\
  &=& 2\pi^2 \sum_{j,k=1}^{g}({\rm Im} \, \Pi)_{jk}^{-1} \,
      \zeta_j(Z)\overline{\zeta_k(Z)} \, 
\frac{{\rm d}Z}{{\rm d}t}\frac{{\rm d}\bZ}{{\rm d}t} 
\eea
using the integral (\ref{intzetbarzet}) and the symmetry of ${\rm Im} \, \Pi$.
By dropping a factor $\half$, we obtain the metric on the one-vortex 
moduli space $\M_1$ (with complex coordinate $Z$)
\be
{\rm d}s^2 = 4\pi^2 \sum_{j,k=1}^{g}({\rm Im} \, \Pi)_{jk}^{-1} \, \zeta_j(Z)
\overline{\zeta_k(Z)} \, {\rm d}Z {\rm d}{\bZ} \,.
\label{Bergman}
\ee

The metric (\ref{Bergman}) is called a Bergman metric on 
${\cal M}_1 \cong \Sigma$ \cite{Lew,Jost}, as we explain in Appendix A. 
Its curvature is negative except, possibly, at a finite number of 
points where it is zero. If we use an orthonormal basis $\eta_j$, 
rather than the canonical basis $\zeta_j$, for the holomorphic 
1-forms, satisfying
\be
\int_\Sigma \eta_j \wedge \overline{\eta_k} = -2i \delta_{jk} \,,
\ee
then the metric becomes
\be
{\rm d}s^2 = 4\pi^2 \sum_{j=1}^{g} 
\eta_j(Z) \overline{\eta_j(Z)} \, {\rm d}Z {\rm d}{\bZ} \,.
\ee
Although this formula looks simpler, it describes the same Bergman 
metric as before, since it is related to (\ref{Bergman}) by 
pull-back (an isometry).

The Bergman metric  (\ref{Bergman}) is also simply the metric (\ref{metricJ}) 
on the Jacobian, restricted to $\M_1$. In fact, the coordinates of the 
image of $P$ in ${\rm Jac}(\Sigma)$ under the Abel--Jacobi map are 
\be
\chi_j(Z) = \int_{Z_0}^Z \zeta_j \,,
\ee
so their differentials are ${\rm d}\chi_j = \zeta_j(Z) \, {\rm d}Z$, and
the restriction of (\ref{metricJ}) to $\M_1$ is 
\be
{\rm d}s^2 = 4\pi^2 \sum_{j,k=1}^{g} ({\rm Im} \, \Pi)_{jk}^{-1} \, \zeta_j(Z)
\overline{\zeta_k(Z)} \, {\rm d}Z {\rm d}{\bZ} \,.
\ee
Equivalently, the Bergman metric is obtained from the flat metric 
(\ref{flatmetric}) by restricting to $\M_1$ using (\ref{c_jJac}).

One of the most interesting properties of $\M_1$ is its total volume
(area in this case). This is
\be
{\rm Vol}(\M_1) = 2\pi^2{\rm  i}\sum_{j,k=1}^{g} 
({\rm Im} \, \Pi)_{jk}^{-1}\int_\Sigma \zeta_j \wedge
\overline{\zeta_k} = 4\pi^2 \sum_{j=1}^{g}I_{jj} = 4\pi^2 g \,,
\label{VolM1}
\ee
where $I$ is the $g\times g$ unit matrix, with trace $g$. Recall that
the ambient space ${\rm Jac}(\Sigma)$ has volume $(2\pi)^{2g}$.

\section{Comparison with Samols' formula}
\news

A general formula for the metric on the $N$-vortex moduli space $\M_N$ 
was established by Samols \cite{Sam,ManSut}. Samols showed, working with 
the linearised vortex equations and the kinetic part of the
lagrangian (\ref{lagran}), that the metric, which is  
defined as an integral over $\Sigma$, can be reduced to a local form,
depending on data at the vortex centres. 
We shall now show that our formula for the metric of one dissolving vortex 
is consistent with the $\tau \rightarrow \frac{4 \pi }{A}$ limit of 
Samols' formula for the metric on $\M_1 \cong \Sigma$.

Let $\phi$ represent the section of ${\mathcal L}\rightarrow \Sigma$ 
in a trivialising patch which, together with the 
connection, solves the vortex equations in unitary gauge for 
one vortex centred at $Z$; this $\phi$ has a simple zero at $Z$. The 
gauge invariant quantity $\log |\phi|^2 - \log |z-Z|^2$ has a 
Taylor expansion around $Z$ \cite{JafTau}
\be
\log |\phi|^2 - \log |z-Z|^2 = a + \half {\bar b} (z-Z) + \half b (\bz - \bZ)
+ \cdots
\ee
where $a,b,\cdots$ depend on $Z$. [N.B. this $a$ is a real function 
of $Z$; it is not the connection.] Then Samols' formula for the 
metric on $\M_1$ (not dividing out by the mass of the vortex) is
\be
{\rm d}s^2 = \pi \left( \tau\Omega + 2 \frac{\pr b}{\pr Z}
\right) \, {\rm d}Z {\rm d}\bZ \,,
\label{Samols}
\ee 
where $\Omega$ and $\frac{\pr b}{\pr Z}$ (which is real) are 
evaluated at $Z$. This formula shows that the metric on $\M_1$ is in the same 
conformal class as the original metric $g_\Sigma$ on $\Sigma$.

We now show that in the dissolving limit, (\ref{Samols}) reduces
to the Bergman metric (\ref{Bergman}). The key is to look at the
complex gauge transformation that converts $\phi$ to holomorphic
gauge. By combining (\ref{unitohol0}) and (\ref{unitohol1}), we see that
\be
{\rm e}^
{\left\{\frac{\tau}{4}\K -{\rm i}\overline{\int_0^z \omega}\right\}} \, \phi 
=: \phi_{\rm hol}
\label{phiholodef}
\ee
is a holomorphic function of $z$ vanishing at $z=Z$. Apart
from its normalisation factor, it depends on $Z$ holomorphically. So
it has an expansion about $Z$,
\be
\phi_{\rm hol}(z) = A(Z,\bZ)\big( z-Z + B(Z)(z-Z)^2 + \cdots \big) \,.
\ee
Also, using (\ref{c_kP}),
\be
\omega = \pi{\rm i}\sum_{j,k=1}^{g}({\rm Im} \, \Pi)_{jk}^{-1}
\left(\overline{\int_{Z_0}^Z \zeta_j}\right) \zeta_k \,,
\ee
so, combining these formulae with (\ref{phiholodef}), we find
\be
\phi = A(Z,\bZ)\big( z-Z + B(Z)(z-Z)^2 + \cdots \big)
{\rm e}^{\left\{ -\frac{\tau}{4}\K + \pi \sum_{j,k=1}^{g}
({\rm Im} \, \Pi)_{jk}^{-1}
\left( \int_{Z_0}^Z \zeta_j \right) \overline{\int_0^z \zeta_k}
\right\}}
\ee
and therefore
\bea
\log|\phi|^2 &-& \log|z-Z|^2 \nonumber \\
&\quad& = - \frac{\tau}{2}\K + \pi \sum_{j,k=1}^{g} ({\rm Im} \, \Pi)_{jk}^{-1}
\left[ \left( \int_{Z_0}^Z \zeta_j \right) \overline{\int_0^z \zeta_k} 
+ \left( \overline{\int_{Z_0}^Z \zeta_j}\right)\int_0^z \zeta_k
\right] \nonumber \\
&\quad& + \log|A(Z,\bZ)|^2 + B(Z)(z-Z) 
+ \overline{B(Z)}(\bz - \bZ) + \cdots \,.
\eea
The Taylor coefficient $\half b$ is the $\bz$-derivative of this, 
evaluated at $Z$,
\be
\half b = - \frac{\tau}{2}\pr_{\bZ}\K
+ \pi  \sum_{j,k=1}^{g}({\rm Im} \, \Pi)_{jk}^{-1} 
\left( \int_{Z_0}^Z \zeta_j \right)
\overline{\zeta_k(Z)} + \overline{B(Z)} \,.
\ee
Here all quantities are functions of $Z$ and $\bZ$. 
Taking the $Z$-derivative, multiplying by $4$, and 
evaluating again at $Z$, gives
\bea
2 \frac{\pr b}{\pr Z} &=& 
- 2 \tau\pr_Z\pr_{\bZ}\K 
+ 4\pi \sum_{j,k=1}^{g}({\rm Im} \, \Pi)_{jk}^{-1}
\zeta_j(Z)\overline{\zeta_k(Z)} \\
      &=& -\tau\Omega(Z,\bZ) 
+ 4\pi \sum_{j,k=1}^{g}({\rm Im} \, \Pi)_{jk}^{-1}
\zeta_j(Z)\overline{\zeta_k(Z)} \,, 
\eea
and hence Samols' formula (\ref{Samols}) reduces to
\be
{\rm d}s^2 =4\pi^2  
\sum_{j,k=1}^{g} ({\rm Im} \, \Pi)_{jk}^{-1} \, \zeta_j(Z)
\overline{\zeta_k(Z)} \, {\rm d}Z {\rm d}{\overline{Z}} \,,
\ee
the Bergman metric. The conformal factor $\Omega$ cancels in this limit.

In general, Samols' formula cannot be used to calculate the vortex 
metric away from the 
dissolving limit, because the coefficient $b$ is not known
explicitly. However, the total volume of the moduli space, 
${\rm Vol}(\M_N)$, is known exactly \cite{ManNas,ManSut}. For 
$N$ vortices on a compact surface of genus $g$, with $N \le g$,
\be
{\rm Vol}(\M_N) = \pi^N \sum_{n=0}^N 
\frac{(4\pi)^n\left(\tau A-{4\pi N}\right)^{N-n} g!}
{n!(N-n)!(g-n)!} \,.
\ee
In the dissolving limit, where $\tau \rightarrow 
\frac{4\pi N}{A}$, the leading term is that with $n=N$, giving
\be
{\rm Vol}(\M_N) = (4\pi^2)^N \frac{g!}{N!(g-N)!} \,,
\ee
with subleading corrections of order $\tau A - 4\pi N$.
For $N=1$ (and $g \ge 1$) this volume is $4\pi^2 g$, agreeing with
(\ref{VolM1}). For $N=g$, the volume in the dissolving limit is
$(4\pi^2)^g$, which is the volume of ${\rm Jac}(\Sigma)$.

Note that for $N=1$, the exact result is
\be
{\rm Vol}(\M_1) = 4\pi^2 g + \pi(\tau A - {4\pi}) \,.
\ee
This shows that the calculation leading to (\ref{VolM1})  
only gives the leading term in ${\rm Vol}(\M_1)$ in the dissolving limit.

\section {Geometry of dissolving multivortices} \label{mvortices}
\news

In this section, we address the problem of understanding the dissolving limit
$\tau \rightarrow \frac{4\pi N}{A}$ for multivortices, assuming
that the two inequalities
\be \label{highg}
1<N<g
\ee
hold. As we have already stated, the image of the moduli space of vortices 
${\cal M}_N \cong {\rm Sym}^N \Sigma$ under the Abel--Jacobi map (well 
defined after a base point $P_0 \in \Sigma$ is chosen) is a subvariety 
$W_N$ of the Jacobian, but this map is in general not an embedding, 
in contrast to the 
$N=1$ case. In analogy with our results in Section~\ref{1vortex}, 
one would expect the relevant metric (in the dissolving limit) to 
be a pull-back of the flat metric on ${\rm Jac}(\Sigma)$ coming from 
the polarisation. However, at critical points of the Abel--Jacobi 
map the pull-back degenerates, as we shall explain.

As before, we define the dissolving limit as the situation where the 
connection ${\rm d}_a$ is projectively flat on a line bundle 
${\mathcal L}\rightarrow \Sigma$ of degree $N$, thus approximating 
the second vortex equation (\ref{Bogo2}) by equation 
(\ref{projflatf}) with $\tau$ close to $\frac{4 \pi N}{A}$. 
The first vortex equation (\ref{Bogo1}) in turn is replaced by 
the constraint that we only consider holomorphic line bundles of 
degree $N$ on $\Sigma$ that are represented by some effective 
divisor --- in other words, line bundles which admit at least one 
(nontrivial) global holomorphic section. This picks out precisely 
the locus $W_N$ inside  ${\rm Jac}(\Sigma)$.

To understand the limiting vortex metric on ${\cal M}_N$, we can follow 
essentially the same steps as in the calculation of 
Section~\ref{1vortex}. Suppose that we have two effective divisors 
of degree $N$,
\be
D_0=\sum_{l=1}^{N}P_l  \qquad \text{and}\qquad D=\sum_{l=1}^N 
Q_l \,,
\ee
where the points $P_l$ and $Q_l$ are
not necessarily distinct. One can pair points in $D_0$ with 
points in $D$ in some arbitrary order, construct a differential 
of the third kind on $\Sigma$ with 
poles of residues $\pm 1$, respectively, on the points of each pair,
and then add all these $N$ differentials to obtain a differential of 
the third kind $\xi^{D_0}_{D}$, uniquely defined up to the addition 
of a differential of the first kind. Setting as before 
$f(z):=-\int_{0}^z \xi_{D}^{D_0}$, where the integration is along a 
path inside $\Sigma_{\rm poly}$ avoiding the points of both $D_0$ and $D$,
we obtain a multivalued function on 
$\Sigma_{\rm poly}\setminus \{ P_1 ,\ldots, P_N,Q_1,\ldots,Q_N\}$ 
which branches logarithmically over the points of 
the divisors; but the object $H^{D_0}_{D}(z):={\rm e}^{f(z)}$ is 
single-valued on $\Sigma_{\rm poly}$, with divisor of zeroes minus divisor 
of poles $(H^{D_0}_{D})=D-D_0$. We can use the 
freedom of adding a holomorphic 1-form to $\xi_{D}^{D_0}$ to make
its $a$-periods vanish, and then $H^{D_0}_{D}$ is defined uniquely 
from the divisors $D_0$ and $D$. 

All of this mimics the
construction in Section~\ref{1vortex} except for one issue: 
over $\Sigma$, it may turn out that $H^{D_0}_D$ is globally defined 
as a meromorphic function (and hence describes a trivial Hecke 
modification, relating two holomorphic sections of the same 
bundle). This occurs precisely when $D_0$ and $D$ are effective 
divisors belonging to the same linear equivalence class. In this
case, the two divisors sit on the same fibre of the Abel--Jacobi map, 
that is, they represent zeroes of two linearly independent, nontrivial 
holomorphic sections of the same holomorphic line bundle over $\Sigma$,
and the extra transition factors ${\rm e}^{\mu_j}$ associated to 
$\xi_{D}^{D_0}$ via (\ref{transitfact}) are trivial. (The factors 
${\rm e}^{\nu_j}$ are already trivial from the normalisation
of $a$-periods.) Using the bilinear relation
\be
\oint_{b_j} \xi^{D_0}_D = -2 \pi {\rm i}\int^{D}_{D_0}\zeta_ j 
:=2 \pi {\rm i} \sum_{l=1}^{N}\left( 
\int_{P_0}^{P_l} \zeta_j - \int_{P_0}^{Q_l} \zeta_j \right) \,,
\ee
where the paths of integration are contained in $\Sigma_{\rm poly}$, 
we can express the transition factors across $a$-edges as
\be
\mu_j = -2 \pi {\rm i}\int_{D_0}^{D} \zeta_j \quad 
({\rm mod} \; 2 \pi {\rm i}) \,.
\ee

Our second point of view of the divisor shift 
also extends to the multivortex case, and the harmonic 1-form that 
describes the change of the connection corresponding
to the divisor shift can be readily computed: 
\be
\alpha = \pi {\rm i}\sum_{j,k=1}^{g}({\rm Im} \, \Pi)_{jk}^{-1}
\left[\left(\overline{\int_{D_0}^D \zeta_j}\right) \zeta_k 
- \left(\int_{D_0}^D \zeta_j\right)\overline{\zeta_k}\right] \,.
\ee
We now regard the divisor $D_0$ as fixed, and the divisor $D$ as
giving the centres of an $N$-vortex solution, i.e. a point of 
${\cal M}_N$. We assign each point $Q_l$ in the divisor $D$ the 
coordinate $z(Q_l)=: Z_l$. Then, to shift the vortices to 
neighbouring points $Q_l + \delta Q_l$, with a corresponding 
shift in coordinates $z(Q_l + \delta Q_l) - z(Q_l) =: \delta Z_l$,
we calculate the infinitesimal change of the harmonic 1-form to be
\be
\delta \alpha = \pi{\rm  i} \sum_{j,k=1}^{g}\sum_{l=1}^{N}
({\rm Im} \, \Pi)_{jk}^{-1}
\left[\overline{\zeta_j(Z_l)} \, \delta\overline{Z_l} \, \zeta_k -
  \zeta_j(Z_l) \,\delta Z_l \, \overline{\zeta_k}\right] \,,
\ee
and this gives rise, when the vortex motion is dynamical, to the electric field
\be
e = \pi{\rm i}\sum_{j,k=1}^{g}\sum_{l=1}^{N} ({\rm Im} \, \Pi)_{jk}^{-1}
\left[\overline{\zeta_j(Z_l)}\frac{{\rm d}\overline{Z_l}}{{\rm d}t} \zeta_k 
- \zeta_j(Z_l)\frac{{\rm d}Z_l}{{\rm d}t} \overline{\zeta_k}\right] \,.
\ee
(We are making the simplifying assumption that the divisor 
$D$ is reduced, so that the coordinates
$Z_l$ are well defined, but this is not essential.)
The kinetic energy associated with the multivortex motion is therefore
\be
T = \half \int_\Sigma e \wedge *e 
  =  2\pi^2 \sum_{j,k=1}^{g}\sum_{l, m=1}^{N}({\rm Im} \, \Pi)_{jk}^{-1} \,
      \zeta_j(Z_l)\overline{\zeta_k(Z_m)} \, 
    \frac{{\rm d}Z_l}{{\rm d}t}\frac{{\rm d}\overline{Z_m}}{{\rm d}t} \,. 
\ee
As for $N=1$, this resembles the kinetic energy for a particle 
moving on the complex manifold ${\cal M}_N = {\rm Sym}^N \Sigma$ with 
local complex coordinates $Z_1,\ldots, Z_N$ (which
degenerate at points where the vortex centres coincide), and whose
metric is the restriction of the flat K\"ahler metric on 
${\rm Jac}(\Sigma)$. The K\"ahler $(1,1)$-form on ${\cal M}_N$ is
given locally by
\be \label{omegadis}
\omega_{\rm diss}= 2 \pi^2 {\rm i}  \sum_{j,k=1}^{g}\sum_{l, m=1}^{N}
({\rm Im} \, \Pi)_{jk}^{-1} \,
 \zeta_j(Z_l)\overline{\zeta_k(Z_m)} \, {{\rm d}Z_l}
\wedge{{\rm d}\overline{Z_m}}  
\ee
and is the pull-back, by the Abel--Jacobi map on degree $N$ divisors, 
of the symplectic structure $\Omega_{\rm J}$ on ${\rm Jac}(\Sigma)$
associated to the polarisation. 

A crucial difference, however, from the $N=1$ case 
is that for $N>1$ this $(1,1)$-form is typically degenerate on some locus 
where the rank drops down. This is because the Abel--Jacobi map typically has
critical points when $N>1$, and one is left with a degenerating metric, 
for which the existence and uniqueness of geodesics associated to 
any point and direction may not hold on this locus. 
(The global 2-form $\omega_{\rm diss}$ is still closed, as
it is the pull-back of the closed 2-form $\Omega_{\rm J}$.) 
So $\omega_{\rm diss}$ only defines a Riemannian structure
over the set of regular points, which is an open subset of 
${\rm Sym}^N \Sigma$. Effective divisors on this subset represent 
line bundles that do not admit independent holomorphic sections (with 
different divisors of zeroes). In contrast, in the language of algebraic 
geometry~\cite{ArCoGrHa}, $\omega_{\rm diss}$ is degenerate over 
``special'' effective divisors, which run or move in nontrivial linear systems.
The directions of degeneracy on ${\rm Sym}^N \Sigma$ are precisely those 
along the complete linear system associated with a special divisor $D$, 
that is, motions from $D$ into effective divisors linearly equivalent 
to $D$. Physically, in the dissolving limit of vortices, motion at 
finite speeds in these directions occurs with no electric field 
and hence no kinetic energy. More usefully, one can say that for 
vortex motion with positive kinetic energy, motion along these 
linear systems could occur infinitely fast in the dissolving limit, 
so that (metrically) these linear systems collapse to points. 

The sets of special divisors $D$, sitting on exceptional fibres of the 
Abel--Jacobi map, are complex projective spaces whose 
dimension $\ell$ can be related to sheaf cohomology via the 
Riemann--Roch theorem~\cite{GriHar}:
\begin{eqnarray}
\ell & =&  \dim_\mathbb{C} \mathbb{P}(H^0(\Sigma,{\cal O}(D)))
\\
&=& \dim_\mathbb{C} H^1(\Sigma,{\cal O}(D))+ \deg D -g+1-1   \\
&=&  \dim_\mathbb{C} H^1(\Sigma,{\cal O}(D)) +N-g \,.
\end{eqnarray}
The divisor $D$ is special precisely when the following strict 
inequality holds:
\be
 \dim_\mathbb{C} H^1(\Sigma,{\cal O}(D)) 
= \dim_\mathbb{C} H^0(\Sigma,{\cal O}(K_\Sigma-D))^* >g-N \,.
\ee

The relations among the geometry of linear systems on $\Sigma$, 
exceptional fibres of the Abel--Jacobi map, and singularities of 
the subvariety $W_N \subset {\rm Jac}(\Sigma)$ are summarised in the 
beautiful Riemann--Kempf theorem, which essentially says that a 
point $w \in W_N$ is a singularity of multiplicity
${g-N+\ell \choose \ell}$, its tangent cone  being the union of 
images of the tangent spaces ${\rm T}_{D}{\rm Sym}^N \Sigma$ by the 
differential of the Abel--Jacobi map, where the effective divisor 
$D$ runs over the complete linear system associated with 
(i.e.\ is in the fibre over) $w$. The subvarieties 
$W_N \subset {\rm Jac}(\Sigma)$ are locally given by determinantal 
equations, and their structure is an important topic in the 
modern algebraic geometry of curves~\cite{ArCoGrHa}.

To illustrate more concretely the behaviour of the Abel--Jacobi map for 
$N>1$ and the structure of its image $W_N$ as a complex $N$-fold inside the Jacobian, 
we briefly describe the possible behaviours at low vortex number.
The qualitative behaviour at a given genus depends crucially on 
the complex structure of $\Sigma$, e.g.\ on whether $\Sigma$ is hyperelliptic, 
and on what kind of linear systems the geometry of $\Sigma$ allows. For more
information, the reader is referred to the textbooks~\cite{ArCoGrHa, Mum}.\\

{\bf Example 1:} 
For $N=2$, the lowest-genus case where (\ref{highg}) is satisfied 
is $g=3$. In this situation there are two subcases.
If $\Sigma$ is a nonhyperelliptic curve (the generic situation), the 
image $W_2 \subset {\rm Jac}(\Sigma)$ of the Abel--Jacobi
map is smooth, and just a copy of the moduli space 
${\cal M}_2 = {\rm Sym}^2 \Sigma$ inside the Jacobian. In fact, this is 
the only case with $N>1$ where the 2-form (\ref{omegadis}) is globally  nondegenerate, and the
dissolving limit metric is regular everywhere. If $g=3$ but $\Sigma$ is 
hyperelliptic, then $W_2$ already has a singularity. 
$W_2$ is the singular complex surface got from the smooth surface
${\rm Sym}^2 \Sigma$ by blowing down a copy of $\mathbb{CP}^1$ to a point, 
which is a double point in $W_2$ \cite{ReiS}. The exceptional $\mathbb{CP}^1$ 
fibre that is blown down is the pencil of degree two divisors 
that are orbits of the hyperelliptic involution (a $g^{1}_{2}$); the space of orbits 
is the quotient of $\Sigma$ by the hyperelliptic involution, which is a 
$\mathbb{CP}^1$ that embeds in ${\rm Sym}^2 \Sigma$ 
holomorphically with noncontractible image. This exceptional fibre 
has an analogue for any moduli space of 2-vortices on a 
hyperelliptic curve $\Sigma$~\cite{BoRo}.\\

{\bf Example 2:}
If $N=3$, the simplest situation requires $g=4$. Then there are three 
subcases. If $\Sigma$ is not hyperelliptic, one can show that it can be 
obtained as an intersection of a quadric $Q$ and a cubic $C$ in 
$\mathbb{CP}^3$. The first subcase is when $Q$ is smooth, hence 
biholomorphic to $\mathbb{CP}^1 \times \mathbb{CP}^1$. Then $C$ 
meets each projective line of the form $\{ P_1\} \times \mathbb{CP}^1$ 
or $\mathbb{CP}^1 \times \{ P_2\}$ in $Q$ at three points, so 
$\Sigma=Q\cap C$ projects to either of the two $\mathbb{CP}^1$ factors 
of $Q$ as a 3-cover. The pre-images of points in $\mathbb{CP}^1$ 
by the two projections form effective divisors of degree $3$ moving 
in two pencils (i.e.\ parametrised by two projective lines), and 
describe two copies $F_1, F_2$ of $\mathbb{CP}^1$ inside ${\rm Sym}^3
\Sigma$, which are $g^{1}_{3}$'s on $\Sigma$ . These are the exceptional fibres of the Abel--Jacobi
map. Physically, there is a metric singularity in the dissolving limit
if the three vortices are centred at the points of these particular
divisors. The image $W_3$ can be obtained by blowing down 
these rational curves $F_1, F_2$ to two points, which are ordinary double points 
of the 3-fold. The second subcase is when $\Sigma$ is not hyperelliptic,
hence $\Sigma=Q\cap C$ as before, but now $Q$ is singular (a quadric 
cone); then $Q$ can be described as a family of projective lines
parametrised by a $\mathbb{CP}^1$ and all meeting at
the singular point. Each line in the family again meets $C$ at three 
points, and so $\Sigma$ inherits one pencil of degree $3$ effective
divisors (a $g^{1}_{3}$), which is the only exceptional fibre of the Abel--Jacobi map.
The image $W_3$ in this case is again got by blowing down this 
$\mathbb{CP}^1$ fibre, and this results in a double point in 
the 3-fold which has higher multiplicity. The third and last subcase
occurs when $\Sigma$ is hyperelliptic. The exceptional fibres here form a 
complex surface inside ${\rm Sym}^3 \Sigma$, namely, the locus of effective
divisors on $\Sigma$ consisting of adding any point of $\Sigma$ to the 
$\mathbb{CP}^1$ of hyperelliptic orbits described in the 
previous example; this can be described as a family of pencils (i.e.\ $g^{1}_{3}$'s)
parametrised by $\Sigma$. Then $W_3$ is obtained from ${\rm Sym}^3 \Sigma$ 
by blowing down this surface to a curve isomorphic to $\Sigma$.

\section {Dissolving multivortices near a singularity} \label{singular}
\news

To understand the behaviour of the geodesic flow close to a 
singularity, we shall now analyse in detail the simplest situation, 
which occurs in the scattering of two vortices on a hyperelliptic 
Riemann surface of genus three.

We start by recalling that the image $W_2$ of the Abel--Jacobi map 
for degree two effective divisors 
\be
{\rm AJ}_2 : {\rm Sym}^2 \Sigma \longrightarrow {\rm Jac} (\Sigma)
\ee
on a hyperelliptic curve $\Sigma$ with $g=3$ has a double point, whose 
blow-up is the exceptional fibre in ${\rm Sym}^2 \Sigma$, which is a 
projective line (see Example~1 in Section~\ref{mvortices}). Since we are only interested in the leading 
local behaviour near this critical locus, we will not need to 
use theta-functions, and will instead take the standard algebraic model
\begin{equation} \label{standsing}
t_3^2 = t_1 t_2
\end{equation}
for the double point, using local coordinates $t_i:U\rightarrow 
\mathbb{C}$ centred at the singularity; so (\ref{standsing})
gives a local equation for the image of $W_2 \cap U \subset 
{\rm Jac}(\Sigma)$ under the coordinate system, which
we may regard as a hypersurface $W'_2$ in an open neighbourhood $U'$ of 
the origin of $\mathbb{C}^3$. Now we blow up $(0,0,0) \in U'$, to obtain
a 3-fold $\widetilde U'$ together with a holomorphic map 
$\pi: \widetilde{U'} \rightarrow U'$ which has
$\pi^{-1}(0,0,0) = \mathbb{P}({\rm T}_{(0,0,0)}U')\cong \mathbb{CP}^2$ 
but is one-to-one everywhere else. We recall how this is 
constructed~\cite{Bal}. The manifold $\widetilde U'$ can be regarded as 
the subset of $U' \times \mathbb{CP}^2$ defined by the incidence relation
\begin{equation} \label{incidence}
t_i v_j = t_j v_i \qquad \text{for all}\quad i,j \in \{1,2,3 \}
\end{equation}
where $v_j$ are homogeneous coordinates on the projectivisation 
$\mathbb{CP}^2$ of the tangent space at the origin, and the map 
$\pi$ is simply the projection ${\rm pr}_{U'}$ onto the first factor. 
In the open set of $U' \times \mathbb{CP}^2$ where $v_3 \ne 0$, for example,
$\widetilde{U}'$ is described by the system of equations
\be
t_1= \frac{v_1}{v_3} t_3, \quad t_2=\frac{v_2}{v_3} t_3
\ee
which has constant rank 2, and this determines a 3-dimensional 
submanifold. Since the incidence relation (\ref{incidence}) is
trivially satisfied for $(t_1,t_2,t_3)=(0,0,0)$, we get indeed the 
whole of the $\mathbb{CP}^2$ factor as exceptional fibre. 

Imposing the equation (\ref{standsing}), we obtain a surface 
$\widetilde W'_2 \cap \widetilde U'$ which is smooth; the singularity is 
replaced by the conic $v_3^2 = v_1v_2$ in the exceptional fibre 
$\mathbb{CP}^2$, which is itself a projective line $\mathbb{CP}^1$, 
and the restriction
\begin{equation} \label{resolution}
\pi |_{\widetilde W'_2\cap \widetilde U' }:  
{\widetilde W'_2\cap \widetilde U'} 
\rightarrow W'_2 \cap U'
\end{equation}
provides a local resolution of the double point on the surface. To 
find the resolution map explicitly, we should use a system of two 
local coordinates where a dense subset of the exceptional fibre is 
visible; for example, an affine coordinate on the $\mathbb{CP}^1$ 
factor, say $q=\frac{v_3}{v_1}$, together with one of the coordinates 
on the first factor, say $p=t_1$. In these coordinates, the projection is
given by
\be
(p,q) \mapsto (t_1,t_2, t_3) = (p,p q^2,  pq ) \in U' \,.
\ee
Working on such local patches, it is not hard to see that the 
projection of $\widetilde W'_2 \cap \widetilde U'$ onto the second factor 
of $\widetilde U' \times \mathbb{CP}^2$ can be understood as a 
restriction of the standard projection
\be \label{cotangent}
{\rm T}^{*(1,0)}\mathbb{CP}^1 \longrightarrow \mathbb{CP}^1
\ee
to a neighbourhood of the (image of the) zero section,
which gives a very concrete picture of the resolution. The exceptional 
fibre of ${\rm AJ}_2$ is identified with the zero section, parametrised by $q$,
and our complex coordinate $p$ parametrises the cotangent fibres.

We want to understand the effect of pulling back a K\"ahler metric 
on $U'$ to the blow-up $\widetilde U'$, and in particular the behaviour 
of the geodesic flow near the exceptional fibre where the metric becomes
degenerate. The K\"ahler metric we consider is the standard euclidean 
metric on $U'$, $g_0=|{\rm d}t_1|^2 + |{\rm d}t_2|^2 
+ |{\rm d}t_3|^2$, as the qualitative behaviour of the flow will 
not depend on anisotropy factors. Pulling back to $\widetilde U'$ we obtain
\begin{eqnarray}
\widetilde g = \pi^{*} g_0&=&(1+|q|^2 + |q|^4){\rm d}p\, {\rm d}\bar p 
+ |p|^2(1+4|q|^2){\rm d}q \,{\rm d}\bar q \nonumber \\
&& + \bar p q (1+2|q|^2) {\rm d}p \,{\rm d}\bar q 
+ p  \bar q (1+2|q|^2) {\rm d}q \,{\rm d}\bar p \,. 
\end{eqnarray}
As expected, this tensor defines a K\"ahler metric in the complement 
of the complex line with equation $p=0$, but its rank 
(over $\mathbb{R}$) drops from 4 to 2 on this line, 
which corresponds to an affine piece of the exceptional 
$\mathbb{CP}^1$ fibre of the Abel--Jacobi map. To understand the 
geodesic flow, we should first compute the Christoffel 
symbols. For a K\"ahler metric this calculation simplifies, and 
moreover Christoffel symbols mixing holomorphic and 
anti-holomorphic directions automatically vanish~\cite{Bal}. We find:
\begin{eqnarray}
&&\widetilde \Gamma^q_{pq}= \widetilde \Gamma^q_{qp} =\frac{1}{p}, 
\qquad \widetilde \Gamma^p_{pq}
=\widetilde \Gamma^p_{qp}=\widetilde \Gamma^q_{pp}=\widetilde 
\Gamma^p_{pp}=0\,,\\
&&\widetilde \Gamma^p_{qq}=-\frac{2 p \bar q^2}{1+4|q|^2+|q|^4}\,, 
\qquad \widetilde 
\Gamma^q_{qq}=\frac{2 \bar q (2+|q|^2)}{1+4|q|^2+|q|^4} \,. 
\end{eqnarray}
These lead to the following geodesic equations:
\begin{equation} \label{geod1}
\ddot p -\frac{2 p \bar q^2 \dot q^2}{1+4|q|^2+|q|^4} =0 \,,
\end{equation}
\begin{equation} \label{geod2}
\ddot q +\frac{2 \dot p \dot q }{p} 
+ \frac{2 \bar q (2+|q|^2) \dot q^2}{1+4|q|^2+|q|^4}=0 \,,
\end{equation}
where the derivatives are with respect to a parameter $s$, say.

An obvious integral of  motion is the kinetic energy of the geodesic 
flow (up to a constant factor),
\be
(1+|q|^2 + |q|^4)|\dot p|^2 + |p|^2(1+4|q|^2)|\dot q|^2 +
 (1+2|q|^2) ( \bar p q \dot p \dot{\bar q} +  p  \bar q \dot{\bar
   p}\dot q) \,. 
\ee
The conservation of this quantity already implies that the motion on 
the exceptional fibre $\mathbb{CP}^1$ (parametrised by the coordinate 
$q$) is suppressed in its tangent directions: as $p\rightarrow 0$, 
all the kinetic energy must be transferred to motion along 
the transverse directions parametrised by the complex coordinate $p$. In particular, 
any geodesic intersecting the exceptional fibre must do so at 
isolated points of the fibre.

To demonstrate that there are indeed geodesics crossing the
exceptional fibre, we note that the geodesic equations above are 
satisfied by the rays of the tangent cone to $W'_2$, i.e.\  paths 
of the form $s\mapsto (p,q)=(c_1 s, c_2)$ for constants $c_1 \in
\mathbb{C}^*$ and $c_2 \in \mathbb{C}$. These correspond to lifts 
of real straight lines on $U'$ towards the singularity, which hit a point 
on the exceptional fibre corresponding to the complex tangent 
direction their velocity represents, and then continue along the 
same real direction. Since the exceptional fibre
is reached in finite time, the metric on the complement of the 
exceptional fibre in $\widetilde{W_2}$ is not complete.

In fact, such straight ray geodesics are the only geodesics reaching 
the exceptional fibre $\mathbb{CP}^1$. To see this, note first that, 
as long as $\dot p$ is not constant, (\ref{geod1}) implies that 
$\dot q$ cannot be zero. 
Dividing equation (\ref{geod2}) by $\dot q$ (assumed to be nonzero)
and extracting the real part of the resulting equation, we obtain
a new differential equation,
\be
\frac{\ddot{q}}{\dot{q}}+\frac{\ddot{\bar{q}}}{\dot{\bar{q}}}+ \frac{2\dot{p}}{p}+\frac{2\dot{\bar{p}}}{\bar{p}}+
\frac{2(2+|q|^2)(\bar{q} \dot{q} + q \dot{\bar{q}})}{1+4|q|^2+|q|^4}=0,
\ee
which can be integrated to conclude that
\be
(1+ 4 |q|^2 + |q|^4) |p|^4 |\dot q|^2
\ee
is another integral of motion. Thus for $p$ to reach zero, 
$\dot q$ would have to blow up, which cannot happen.
Initial conditions that try to reach the exceptional fibre with 
initial velocities having nontrivial tangent component along 
the $\mathbb{CP}^1$ will be forced to flow rapidly around this 
2-sphere as they approach it transversely.

In terms of vortex motion, the effect of the singularity is that
motion along the special linear system is suppressed. So whenever two
vortices reach points on the surface that are related by the
hyperelliptic involution, they will be unable to move to neighbouring
pairs of points that are also related by the involution. In
particular, it will be impossible to make vortices collide 
head-on onto a Weierstra\ss\  point of the
surface: these are precisely the branch points of the two-fold
holomorphic branched cover $\Sigma \rightarrow \mathbb{CP}^1$, and geodesics 
through them are tangentially preserved by the hyperelliptic
involution near the branch point.

\appendix
\section{On the notion of Bergman metric}
\numberwithin{equation}{section}
\news 

Here we give a brief account of the notion of Bergman metric, around 
which there is occasionally some confusion in the literature. The 
name is always used to refer to a metric on a complex manifold that 
is invariant under local holomorphic maps, but it may
refer to distinct objects, as we shall now review. 

For most authors, the Bergman metric arises in the theory of several 
complex variables and is defined on complex domains using the 
(intrinsic) Bergman kernel that we now describe~\cite{Kra}.
Let ${\cal D}$ be a domain in $\mathbb{C}^n$, i.e.\ a connected 
open subset, where global complex coordinates $z=(z_1,\ldots, z_n)$ 
are defined. The Hilbert space of square-integrable functions
on ${\cal D}$ (with the $L^2$ inner product 
\be \label{L2}
\langle f_1,f_2 \rangle_{L^2} := 
\int_{\cal D} f_1(z) \overline{f_2(z)} \, {\rm d} \mu(z)
\ee
coming from the Lebesgue measure $\mu$ on $\mathbb{C}^n$) has a
distinguished subspace, namely, the set of $L^2$-functions $f$ that 
are holomorphic. This subspace is closed and thus inherits
a Hilbert space structure from~(\ref{L2}).
For each point $z \in \mathbb{C}^n$, the evaluation map
\be
{\rm ev}_z: f \mapsto f(z) \in \mathbb{C}
\ee
is continuous and linear, so the Riesz representation theorem implies 
that this map can be described by an integral operator:
\be
{\rm ev}_z(f)=\int_{\cal D} f(z) K(z,z') \, {\rm d}\mu(z')\,.
\ee
The Bergman kernel is the function of $2n$ complex variables $K$. 
If we let $u_j$ denote elements of an orthonormal basis of the space 
of holomorphic $L^2$-functions, indexed by the integers $j\in \mathbb{N}$, 
one can show~\cite{Kra} that
\be \label{BerKer}
K(z,z')=\sum_{j=1}^{\infty} u_j(z) \overline{u_j (z')} \,.
\ee
So $K$ is holomorphic in the first $n$ and antiholomorphic in the 
last $n$ variables, and to emphasise this we shall denote it as 
$K(z,\bar{z}')$ henceforth.

If we restrict the Bergman kernel to the diagonal and set
\be
{\rm d}s_{(1)}^2=\sum_{k,l=1}^{n}\frac{\partial^2 \log K (z,\bz)}
{\partial z_k \partial \overline{z_l}}\, {\rm d}z_k {\rm d}\overline{z_l} \,,
\ee
we obtain a K\"ahler metric on ${\cal D}$, which is easily seen to 
be invariant under holomorphic maps.
However, if $n=1$ we could also have set (writing $z$ instead of $z_1$)
\be \label{Ber2}
{\rm d}s_{(2)}^2= K (z,\bz) \,{\rm d}z\,{\rm d}{\bz}
\ee
to obtain another K\"ahler metric invariant under holomorphic maps. 
Notice that there is no analogue of ${\rm d}s^2_{(2)}$ if $n>1$. 
Standard textbooks~\cite{Kra} call ${\rm d}s^2_{(1)}$ the Bergman 
metric of ${\cal D}$, but this designation is also used for 
${\rm d}s^2_{(2)}$, which in Bergman's most famous book~\cite{Ber} 
is simply called `the invariant metric'. Notice that the K\"ahler 
form $\omega_{(1)}$ of ${\rm d}s^2_{(1)}$ turns out to be 
$\omega_{(1)} = - \rho_{(2)}$, that is, minus the Ricci form of the K\"ahler 
structure associated to ${\rm d}s^2_{(2)}$, cf.~\cite{Kob}.

Now suppose that instead of a domain ${\cal D} \subset \mathbb{C}$ 
we have a closed Riemann surface $\Sigma$. The space of square-integrable 
functions on $\Sigma$ is very large, but the only ones that are 
holomorphic are the constants. So in this case the above notions 
of Bergman metric (depending on the Bergman kernel as defined) are 
vacuous. The way out is to work with holomorphic 1-forms instead of 
holomorphic functions. If $\Sigma$ has genus $g>1$, the space $H^0(\Sigma,K_\Sigma)$ of
global holomorphic 1-forms on $\Sigma$ is nontrivial and has complex 
dimension $g$. Choosing a basis $\zeta_1, \dots, \zeta_g$, we can set
\be \label{Ber3}
{\rm d}s_{(3)}^2 = \sum_{j=1}^{g} \zeta_j \overline{\zeta_j} 
\ee
and obtain a metric on $\Sigma$. A metric of this type is also sometimes 
referred to as a Bergman metric~\cite{Jost, Kob}, and this is the 
notion that we use in this paper. (These metrics can be extended to 
the study of smooth curves in positive characteristic, and they are
the starting point for Arakelov geometry~\cite{Fal}.) Notice that the 
basis of $H^0(\Sigma,K_\Sigma)$ can be changed by a linear transformation 
$S \in {\rm GL}_g (\mathbb{C})$, and then a new metric will be obtained
whose coefficients with respect to the original basis are related by 
the hermitian matrix $S S^*$. So the space of Bergman metrics of 
type (\ref{Ber3}) has real dimension $g^2$, even though there is twice as
much freedom to choose a basis of holomorphic 1-forms on a compact 
Riemann surface of genus $g$. Each such metric is associated to a 
basis of $H^0(\Sigma,K_\Sigma)\cong \mathbb{C}^g$, or alternatively to a 
hermitian inner product on this complex vector space. One can show 
that all these metrics have nonpositive curvature, and the curvature vanishes
precisely at the Weierstra\ss\ points of $\Sigma$ in the case where $\Sigma$ 
is hyperelliptic.

If $z$ is a local coordinate, one has $\zeta_j = \zeta_j(z){\rm d}z$ for
some local holomorphic function $z \mapsto \zeta_j(z)$ for each $j$, 
and then
\be \label{localBer3}
{\rm d}s_{(3)}^2 = \sum_{j=1}^{g} \zeta_j(z) \overline{\zeta_j(z)} 
\, {\rm d}z \,{\rm d}\bz \,.
\ee
Comparing (\ref{localBer3}) and (\ref{Ber2}), one sees that the 
Bergman kernel (\ref{BerKer}), which does not exist on $\Sigma$, is being replaced
by the quantity $\sum_{j=1}^{g} \zeta_j(z) \overline{\zeta_j(z)}$. 
Notice that we could use this ersatz Bergman kernel to define another 
metric on $\Sigma$ mimicking ${\rm d}s^2_{(1)}$ above in most cases, but 
this would not be a genuine metric everywhere if $\Sigma$ were
hyperelliptic, by our observation above about vanishing curvature.

The arbitrariness in the Bergman metrics (\ref{Ber3}) can be avoided 
if we orthonormalise the holomorphic 1-forms $\zeta_j$ with respect 
to the inner product (\ref{1formsL2}). This amounts to restricting to one
particular metric of type (\ref{Ber3}), namely, the one associated to 
the polarisation of the Jacobian, which defines a hermitian inner 
product on $H^0(\Sigma,K_\Sigma)$ by duality. This is precisely the Bergman 
metric (\ref{Bergman}) that we have shown to describe the motion 
of one vortex on $\Sigma$ in the dissolving limit.

\section*{\large Acknowledgements}

NSM is grateful to the Department of Physics, University of Pisa, for
hospitality, and to the Department of Mathematical Sciences,
Durham University, for hospitality and for a Blaise Pascal award. 
The work of NMR was partially supported by CTQM, University of Aarhus,
and by the European Commission in the framework of the Marie Curie
project MTKD-CT-2006-042360; he is grateful to the University
of Cambridge for hospitality when this collaboration was initiated, and to 
\.{Z}ywomir Dinew and Andrew Swann for useful discussions.

\end{document}